\documentclass[12pt,thmsa,notitlepage]{report}
\usepackage{amssymb}
\usepackage{pppbib}
\usepackage{captions}
\usepackage{graphicx}
\begin{document}
\title{Astrophysical Aspects of Quark-Gluon Plasma}
\author{Daniel Enstr\"{o}m\footnote{daniele@mt.luth.se}
\\ Department of Physics \\
Lule\aa \ University of Technology\\
SE-971 87 Lule\aa , Sweden}
\maketitle
\thispagestyle{empty}
\begin{abstract}
This M.Sc. thesis in Engineering Physics is an overview of the present theory of quark-gluon
plasma (QGP) as well as an analysis of the stability criterion for
possible stable cosmic QGP objects left over from the quark-hadron
transition in the early Universe. It covers fundamental ideas of the formation
and decay of the plasma, including the standard model, QCD, and the
MIT bag model. I discuss the equation of state of a QGP and the
possible signatures for a plasma created in heavy-ion collisions.
Astrophysical aspects of QGP are put forward, including compact
stars and the quark-hadron transition in the early Universe. The
possible role of QGP objects as cosmic dark matter is mentioned.
The analytic part is an investigation of possible stability among cosmic QGP
objects from the early Universe. A model is suggested where a
pressure balance makes a QGP stable against
gravitational contraction and hadronization. The mass/radius
relationship for stability also forbids a direct gravitational
collapse. Finally, the possibility of stable cosmic QGP objects is critically
discussed.
\end{abstract}
\newpage
\vspace*{7cm}
\begin{displaymath}
\mbox{{\it To Ulli, Oskar and Olof}}
\end{displaymath}
\newpage
\thispagestyle{empty}
\tableofcontents
\thispagestyle{empty}
\newpage {\LARGE \textbf{Introduction}}

\pagestyle{plain}

\bigskip
\addcontentsline{toc}{chapter}{Introduction}

This thesis is devoted to the subject of quark-gluon plasma, QGP, a
fairly new branch of high-energy physics. A lot of effort has been
put into the area, primarily into experimental detection of QGP
production in heavy-ion collisions, but as of today, no undisputed
signal seen.

\bigskip

The largest experiments up to date in the search for a QGP is at
the SPS accelerator at CERN in Geneva, where physicists have been looking for a QGP for several years
without success. These experiments are still running, and will
hopefully be continued in one form or the other when the new large
hadron collider, (LHC), currently under construction at CERN, will
be taken into use around the year 2005.

\bigskip

In the US, a major effort to produce and detect a QGP is underway
at the Brookhaven National Laboratory, where a new relativistic
heavy-ion collider, (RHIC) is being built for a planned operation by 1999.

\bigskip

This thesis is not going into details about the experimental
efforts to detect a QGP. I will instead focus on the theory of a
QGP together with an excursion into the exciting field of
astrophysics. The thesis is split up into two parts. One is an
overview of the physics of a QGP, including a short introduction to
quantum chromodynamics (QCD) and the so-called MIT bag model,
 and the other describes the results I have achieved about some astrophysical aspects of
QGP.

\bigskip

In the overview I have tried to include those aspects of QGP that
are widely stated as fundamental and well established. It starts
with a short overview of the standard model and describes the basic
properties of quarks and gluons. The gauge theory (QCD) describing
the strong interaction is given some attention as well as the
QCD-inspired MIT bag model of quarks confined in hadrons. Since the
confinement mechanism is crucial in the phase transition from
hadronic matter to QGP, a rather extensive description is given, as
well as a discussion of the phase transition itself.

\bigskip

The equation of state of a QGP is given, i.e., the relation between
the pressure and temperature inside the plasma. This is followed by
a discussion of the formation and decay of a QGP, which leads into
a subsequent overview of possible experimental QGP signals.

\bigskip

Various astrophysical aspects of QGP are discussed, and the role of
QGP in the phase transitions in the early Universe is mentioned, as
well as the current theory governing compact stars, i.e., neutron,
quark and hybrid stars. This provides a link between the QGP
overview and my own calculations and results.

\bigskip

I examine the possibility of stable cosmic QGP objects surviving
from the quark-hadron transition in the early Universe. I show,
within the theory of general relativity and the equation of state
given by QCD, that stable configurations can occur when the size
and the mass of the QGP have a certain relationship derived from
the so-called Tolman-Oppenheimer-Volkoff equation.

\bigskip

At the end I discuss some general aspects of QGP and point out
certain discrepancies between the model emerging from heavy-ion
collisions and ideas applied to astrophysical QGPs.

\bigskip

In two appendices, chiral symmetry and the physics of phase
transitions are discussed.

\chapter{Quarks and gluons}

\label{chapquark}

\section{The standard model}

Most particle physicists today believe that the standard model of
elementary particle physics more or less describes the fundamental
building blocks of matter along with their interactions (apart from
gravity). It comprises the theories of the electroweak and the
strong interactions. The standard model tells us that we have two
groups of elementary particles, leptons and quarks. In addition,
the different types of interactions included in the model are due
to exchange particles, in the form of vector bosons.
\[
\begin{tabular}{cc}
\hline\hline
\textbf{Quarks} & \textbf{Leptons} \\ \hline\hline
\begin{tabular}{lcc}
Flavour & Mass ($GeV/c^2$) & Charge \\ \hline
\multicolumn{1}{c}{$u$} & $0.3$ & $2e/3$ \\
\multicolumn{1}{c}{$d$} & $0.3$ & -$e/3$ \\
\multicolumn{1}{c}{$c$} & $1.0\rightarrow 1.6$ & $2e/3$ \\
\multicolumn{1}{c}{$s$} & $0.45$ & -$e/3$ \\
\multicolumn{1}{c}{$t$} & $180\pm 12$ & $2e/3$ \\
\multicolumn{1}{c}{$b$} & $4.1\rightarrow 4.5$ & -$e/3$%
\end{tabular}
&
\begin{tabular}{ccc}
Flavour & Mass ($GeV/c^2$) & Charge \\ \hline $\nu _e$ & $<8\times
10^{-9}$ & 0 \\ $e$ & $5.110\times 10^{-4}$ & -$e$ \\ $\nu _\mu $ &
$<2.7\times 10^{-4}$ & $0$ \\ $\mu $ & $0.1057$ & -$e$ \\ $\nu
_\tau $ & $<0.035$ & $0$ \\
$\tau $ & $1.777$ & -$e$%
\end{tabular}
\\ \hline\hline
\end{tabular}
\]

\bigskip

The leptons can only interact by the electroweak interaction,
i.e., the unified electromagnetic and weak interaction. They do not feel
the strong force mediated by the gluons. The quarks, on the other
hand, interact strongly, weakly and electromagnetically.

Leptons and quarks obey certain empirical particle-number
conservation laws. If the neutrinos are massless, one could speak
of conservation of lepton type, i.e., conserved electron $(\nu
_e,e)$, muon ($\nu_\mu ,\mu )$ and tau $ (\nu _\tau ,\tau )$ numbers.

\[
\begin{tabular}{cccc}
\hline\hline
\textbf{Particle} & \textbf{L}$_e$ & \textbf{L}$_\mu $ & \textbf{L}$_\tau $
\\ \hline\hline
$\nu _e,e^{-}$ & 1 & 0 & 0 \\ $\nu _\mu ,\mu^{-}$ & 0 & 1 & 0 \\
$\nu_\tau ,\tau^{-}$ & 0 & 0 & 1 \\ $non-leptons$ & 0 & 0 & 0 \\
\hline\hline
\end{tabular}
\]

\bigskip

All leptons have antiparticles with opposite electric charges and
lepton numbers.

\bigskip

Quark flavour is conserved by the strong interaction, but not by
the weak interaction. One can speak of a total quark number due to
the stability of the proton. Up to this date, experiments and
observations are consistent with total conservation of overall
quark number and of lepton type, but these conservation laws are
not a consequence of any known dynamical principle.

\[
\begin{tabular}{ccccccc}
\hline\hline
\textbf{Quark quantum numbers} & $\mathbf{u}$ & $\mathbf{d}$ & $\mathbf{c}$
& $\mathbf{s}$ & $\mathbf{t}$ & $\mathbf{b}$ \\ \hline\hline
\textit{I}$_3$\textit{\ - isospin 3-component} & 1/2 & -1/2 & 0 & 0 & 0 & 0
\\
\textit{S - strangeness} & 0 & 0 & 0 & -1 & 0 & 0 \\
\textit{C - charm} & 0 & 0 & 1 & 0 & 0 & 0 \\
\textit{B - bottomness} & 0 & 0 & 0 & 0 & 0 & -1 \\
\textit{T - topness} & 0 & 0 & 0 & 0 & 1 & 0 \\ \hline\hline
\end{tabular}
\]

\smallskip

All quarks have their antiparticles, the antiquarks. They have the
same quark quantum numbers and electromagnetical charges apart from
a minus sign.

\bigskip

The three types of interactions included in the model are all
mediated by exchange of vector bosons. The mediator of the
electromagnetic force is the photon, those of the weak force are
the W$^{\pm }$ and the Z$^0$, and the strong force is mediated by
the gluons. The photon carries no electric charge and there is no
direct interaction between photons. The gluons actually carry
colour charge, a sort of ''strong'' charge, and therefore interact
with each other. The gluons and the photons are presumably
massless, but the W$^{\pm }$ and the Z$^0$ are quite heavy.
Therefore, the range of the weak interaction is very short, about
10$^{-18}$ m. The strong interaction has a limited effective range
of $10^{-15}$ m due to colour-screening effects and gluon
self-interaction.
\[
\begin{tabular}{cc}
\hline\hline
\textbf{Particle} & \textbf{Mass\ (}$\mathbf{GeV/c}^2$\textbf{)} \\
\hline\hline
W$^{\pm }$ & $80.33\pm 0.15$ \\ Z$_{}^0$ & $91.187\pm 0.007$ \\
$\gamma $ & $<6\times 10^{-25}$ \\ \hline\hline
\end{tabular}
\]

\section{Quarks}

Experimentally, one has found evidence for six quarks; up $(u)$,
down ($d$), charm ($c$), strange ($s$), top ($t$) and bottom ($b$),
which all are fermions. Their masses differ much, from the up with
a mass of around $300$ MeV/c$^{2}$ to the top-quark with a mass
around $180$ GeV/c$^{2}$\cite{topquark}. All quarks have electric
charges; $+\frac 23e$ for the $u$,$c$ and $t$ and $
-\frac 13e$ for the $d$,$s$ and $b$. Each quark is said to represent a separate flavour.

\bigskip

Quarks, as fermions, obey the Pauli principle, which presented some
major difficulties when the $\Delta ^{++}$ resonance was
discovered. The $\Delta
^{++}$ resonance is a spin/parity J$^P=\frac {3}{2}^{+}$ particle consisting of three up
quarks with parallel spins:
\[
|\Delta ^{++}>=|u^{\uparrow }u^{\uparrow }u^{\uparrow }>.
\]

This situation is not favourable in any way. Three identical
particles with parallel spins violate the Pauli principle. This
problem was solved by introducing a colour charge \cite{color ref}.
The colour quantum number can take three basic values, e.g. "red",
"green" and "blue". Anti-particles can be "antired", "antigreen"
and "antiblue". The Pauli principle is now obeyed, provided that
the three quarks in a baryon have different colours. The $\Delta
^{++}$ resonance has an antisymmetric wave function:

\bigskip
\[
|\Delta ^{++}>=|u_r^{\uparrow }u_g^{\uparrow }u_b^{\uparrow }>.
\]

When the quark concept was invented in the mid 1960s the
physicists were able to categorize and describe most that had been
discovered in accelerator experiments. The particles that are made
of quarks were called hadrons; strongly interacting particles.
Hadrons with half-integer spin were called baryons and hadrons with
integer spin mesons.

\bigskip

The proton and the neutron are two examples of hadrons. Each
consists of three quarks,

\[
|p>=|uud>\hspace{2cm}|n>=|udd>,
\]
and have electric charge $+e$ and $0$, respectively. In the
proton/neutron case, the three quarks that give the nucleon its
properties are called valence quarks. Virtual quark-antiquark pairs
are continuously created and annihilated inside the nucleon.

\bigskip

All in all, the quark family consists of six quarks, with different
flavours, and they all carry electric charge as well as colour
charge.

\bigskip

No free quark has ever been detected, in spite of several extended
searches. Therefore one believes that quarks are confined inside
hadrons, and that the strong force potential between the quarks
increases as the distance gets larger. One would then need an
infinite amount of energy to separate two quarks from each other.
Hence hadrons are colour neutral objects. The quarks forming a
hadron must have a colour combination rendering a colourless,
"white" particle. For instance, the spin-0 meson $\pi ^{+}$ has
three possible colour configurations.

\begin{figure}[h]
\begin{center}
\includegraphics{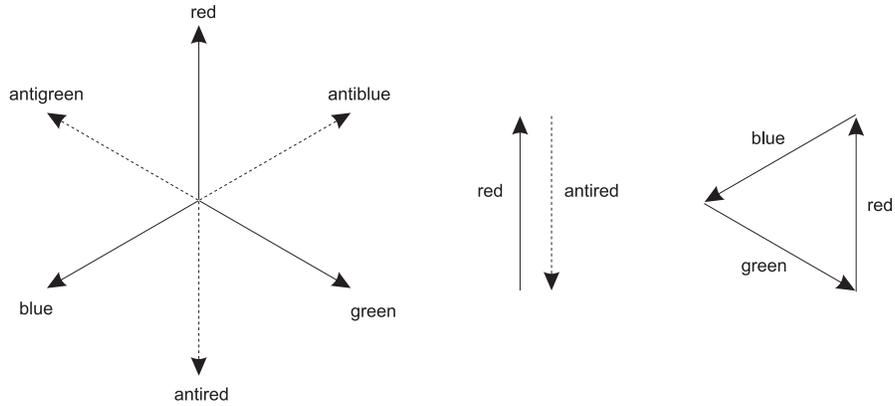}
\end{center}
\caption{\textit{Graphical picture of the relation between the different
colours and two examples of colourless, white states. }}
\end{figure}

\[
|\pi ^{+}>=\left\{
\begin{array}{c}
\begin{tabular}{l}
$u_r\overline{d}_{\overline{r}}>$ \\
$u_b\overline{d}_{\overline{b}}>$ \\
$u_g\overline{d}_{\overline{g}}>$
\end{tabular}
\end{array}
\right.
\]
A physical $\pi ^{+}$ is equal to a quantum-mechanical mixture of
these states.

\bigskip

\section{Gluons}

\bigskip

The force that binds quarks together is the strong interaction. It
is mediated by its exchange particles, the gluons. The gluon is a
spin-1 particle with no mass, and hence the strong interaction
would have a long range would it not be screened. The effective
range of the force is of the order of $10^{-15}$ m. The gluons
carry both colour and anticolour. Since $3\times 3$ colour
combinations exist, the gluons form two multiplets of states, an
octet and a singlet. It is possible to construct all colour states
from the octet and therefore there are eight different gluons. The
ninth state is the totally symmetric state
$R\overline{R}+B\overline{B}+G%
\overline{G}$, which is colourless, and therefore plays no role in
the strong interaction.

\bigskip

Since gluons themselves carry colour they can interact with each
other. This is the main difference between the strong and the
electromagnetic interaction, since photons carry no electric
charge. Gluons can be emitted and absorbed, created and
annihilated, when a strong interaction is involved in a process.

\begin{figure}[h]
\begin{center}
\includegraphics{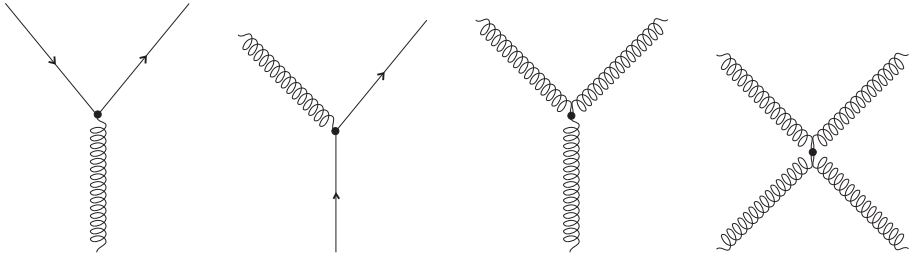}
\end{center}
\caption{\textit{The fundamental interaction Feynman diagrams of the strong
interaction. From left to right, splitting of a gluon into a
quark-antiquark pair, emission of a gluon, and two gluon
self-couplings. }}
\end{figure}

\bigskip

Since quarks make up the proton and the neutron, gluons mediates
the force that keep the proton and the neutron together. The strong
nuclear force that makes the nucleons in a nucleus stick together,
is in quark-gluon terms a leakage
 of the forces that keep the nucleons together, e.g. in the form
 of quark-antiquark pairs.

\bigskip

Since all quarks have the same possibility to carry a certain
colour regardless of flavour, the strong interaction is flavour
independent, i.e., all sorts of quarks have identical strong
interactions.

\bigskip

One consequence of the properties of the strong interaction is that
the strength of the strong force decreases when the quarks are
close to each other. This property gives the quarks what is called
asymptotic freedom at small inter-quark distances.

\bigskip

The strength of the strong interaction is described by the strong
"running" coupling constant $\alpha_s$. It has a direct dependence
on the squared four-momentum transfer in the quark process:

\begin{equation}
\alpha _s(Q^2)=\frac{12\pi }{(33-2N_f)ln(Q^2/\Lambda ^2)}.
\label{strong force factor}
\end{equation}
N$_f$ is the number of quark flavours involved and $\Lambda $ is a
scaling parameter, $\Lambda =0.2\pm 0.1$ GeV. From this expression,
it is easy to see how $\alpha _s$ decreases with increasing
momentum transfer. This is equivalent to decreasing $\alpha_s$ when
the distance to a quark decreases. Hence, asymptotic freedom and
colour confinement are implied by the expression for the strength
of the interaction.

\bigskip

A more detailed account of the quantum field theory of the strong
interaction (QCD) is given in the next chapter.

\bigskip

\chapter{Quantum chromodynamics}

\label{chapqcd} \bigskip

\section{The concept of colour}

\bigskip

A new quantum-mechanical gauge theory of quark interaction was born
in the 1960s and 1970s. It was called QCD, in analogy with quantum
electrodynamics (QED).

\bigskip

The basic idea with QCD is invariance against arbitrary rotations
in colour space. Since the complex rotations of an arbitrary vector
in three-dimensional space are described by unitary $3\times 3$
matrices of unit determinant, the symmetry group of the gauge
transformation is $SU(3)$.

\bigskip

When the concept of colour was invented, the main reason was to
make the quarks obey the Pauli principle, but at that time there
was no evidence that there are exactly three colours. The
experimental verification of the number of colours came in the
beginning of the 1980s \cite{PDG82}. The ratio of the production of
hadrons from $e^{+}e^{-}$ collisions to that of muon pairs reveals
an interesting fact. This is based on an assumption that the
production of hadrons proceeds through creation of a
quark-antiquark pair, which subsequently fragments into hadrons. In
that case, the ratio of the cross-sections can be related to the
sum over the square of the electromagnetical charges of all sorts
of quarks that can be created. Defining $N_c$ as the number of
colours, one gets

\begin{equation}
R=\frac{\sigma (e^{+}e^{-}\rightarrow hadrons)}{\sigma
(e^{+}e^{-}\rightarrow \mu ^{+}\mu
^{-})}=N_c\sum_ie_i^2=\frac{11}9N_c
\label{ratio}
\end{equation}
for i = $u,d,s,c,b$ quark types. The experiments yield $N_c=3$
without doubt \cite{PDG82}.

\bigskip

\section{The QCD Lagrangian}

\bigskip

The principle behind the QCD\ Lagrangian is the demand for
invariance under local colour rotations, i.e., the gauge matrix $U$
changes from one point in space to another. In quantum
electrodynamics (QED) the governing principle of gauge
transformation is changes in the phase of the wave function, as
described by the one-dimensional $U(1)$ symmetry group.

\bigskip

The approach in deriving the QCD Lagrangian is the same as in QED
modified to fit the gauge invariance of the strong interaction. The
fundamental QED Lagrangian is (see, e.g. \cite{QGP}):

\begin{equation}
L_{QED}=i\overline{\psi }\gamma ^\mu (\partial _\mu +ieA_\mu )\psi -m%
\overline{\psi }\psi -\frac 14F^{\mu \nu }F_{\mu \nu },
\label{QED Lagrangian}
\end{equation}
where $F^{\mu \nu }=\partial _\mu A_\nu -\partial {_\nu }A_\mu $ is
the electromagnetic field strength tensor, and $A_{\mu}$ is the
electromagnetic vector field.

\bigskip

Due to the colour property, the wave function of a quark has three
components in colour space, $\psi =(\psi _r,\psi _g,\psi _b)$. A
colour
gauge transformation, $\psi \rightarrow U\psi $, is described by a unitary $%
3\times 3$ matrix $U$ with $det(U)=1$. $U$ can be written as the
imaginary
exponential of a Hermitian matrix $L$, $U=exp(iL)$, where $L^{*}=L$ and $%
tr(L)=0$. All traceless Hermitian $3\times 3$ matrices can be
expressed as
linear combinations of the eight $\lambda $-matrices \cite{GM62}

\begin{equation}
L=\frac 12\sum_{a=1}^8\theta _a\lambda _a . \label{Gell-Mann
non-space}
\end{equation}
Here
\begin{equation}
\left[ \lambda _a,\lambda _b\right] =2if_{abc}\lambda _c\hspace{0.5cm}%
\mbox{and}\hspace{0.5 cm}\left[ \lambda _a,\lambda _b\right] _{+}=\frac
43\delta _{ab}+2d_{abc}\lambda _c. \label{Gell-Mann commutations}
\end{equation}
These so-called Gell-Mann matrices, or $\frac 12\lambda _a$ to be
more
exact, are the eight generators of the Lie group $SU(3)$, and $f_{abc}$ and $%
d_{abc}$ are the antisymmetric and symmetric structure constants.
To make the rotation $U$ space-dependent the real parameter $\theta
_a$ must vary in space, $\theta _a=\theta _a(x)$. This leads to

\begin{equation}
U(x)=exp \left[ \frac{1}{2}\sum_{a=1}^8\theta_{a}(x)\lambda_{a}
\right]
\label{Gell-Mann space dep}
\end{equation}

Thus $\psi \rightarrow U(x)\psi $ gives

\begin{equation}
\partial _\mu [U(x)\psi]=U\partial _\mu \psi +(\partial _\mu U)\psi
=U[\partial _\mu \psi +U^{*}(\partial _\mu U)\psi].
\label{non cov derivative}
\end{equation}
Since $U\partial_{\mu}\psi$ should be present in the right-hand
side one has to introduce a colour potential, the analogue to the
electromagnetic potential $A_\mu $ in QED. That potential is called
$\widehat{A }_\mu $; a $ 3\times 3$ Hermitian matrix. This
$\widehat{A}_\mu $ can be represented as a linear combination of
Gell-Mann matrices:

\begin{equation}
\widehat{A}_\mu (x)=\frac 12\sum_{a=1}^8A_\mu ^a(x)\lambda _a.
\label{Yang-Mills decomposition}
\end{equation}
The field $\widehat{A}_\mu (x)$ is called a Yang-Mills field \cite{Yang-Mill}%
, and if $\widehat{A}_\mu (x)$ changes during a colour rotation
according to

\begin{equation}
\widehat{A}_\mu \rightarrow U^{*}\widehat{A}_\mu U-i\frac 1gU^{*}(\partial
_\mu U),  \label{Yang-Mills transformation}
\end{equation}
the derivative $(\partial _\mu -ig\widehat{A}_\mu )$ will be
invariant under such gauge transformations. Here $g$ is the
so-called strength factor of the strong field. Since the basic
relation is the Dirac equation one can choose a potential that
leaves the derivative invariant under such a gauge transformation
making the whole equation invariant. The replacement
defines a covariant derivative, $D_\mu=\partial _\mu \rightarrow $ $%
(\partial_\mu -ig\widehat{A}_\mu )$.

\bigskip

The next step is to add a kinetic term to the Lagrangian, and a
first guess is a QED-like term, $-\frac 14F^{\mu \nu }F_{\mu \nu
}$. It turns out that one has to modify the definition of the field
strength tensor $F_{\mu \nu }$ in order to keep the theory gauge
invariant. Since no interactions take place between two photons,
QED is an abelian theory and the generators of the $U(1)$ group
commute. QCD, on the other hand, is a non-abelian $SU(3)$ local
gauge theory. The generators are non-commutative, due to the fact
that interactions take place between the gluons. The field strength
tensor therefore changes to

\begin{equation}
F_{\mu \nu }^a=\partial _\mu A_\nu ^a-\partial _\nu A_\mu
^a+gf_{abc}A_\mu
^bA_\nu ^c,  \label{QCD field strength tensor}
\end{equation}
and this definition makes it form invariant under a local colour
gauge transformation, $F_{\mu \nu }\rightarrow U^{*}F_{\mu \nu }U$.

\bigskip

The complete QCD Lagrangian is then

\begin{equation}
L_{QCD}=i\overline{\psi }\gamma ^\mu (\partial _\mu
-ig\widehat{A}_\mu )\psi
-m\overline{\psi }\psi -\frac 14F_{\mu \nu }^aF_a^{\mu \nu }.
\label{QCD Lagrangian}
\end{equation}
Notice the resemblance with the QED Lagrangian, eq. (\ref{QED
Lagrangian}). All the difference lies in the self-interaction term
in the definition of $F_{\mu
\nu }$ and in the exchange of $+ieA_\mu \rightarrow -ig\widehat{A}_\mu$.

\bigskip

\chapter{The confinement of coloured particles}

\label{chapconf}

\section{The confinement mechanism}

The non-observation of single quarks led the physicists to
postulate that no particle can exist in a coloured state, and as a
consequence, to establish the concept of quark confinement.

\bigskip

QCD has complex field equations that make it unsuitable for "exact"
calculations. This leads to extensive use of perturbative methods.
These are, on the other hand, valid only in certain regimes of the
perturbative expansion parameter. In QCD, this parameter is the
strong coupling constant, $\alpha _s$. It gets small when the
four-momentum transfer is large or, equivalently, the distance
involved is small as illustrated by eq. (\ref{strong force
factor}). That makes perturbative QCD valid only in processes that
have these characteristics. This is unfortunate because the
distance when $\alpha_{s}$ gets too large for a perturbative
expansion, is about the same as the dimensions of a hadron.
Therefore, the confinement mechanism cannot be derived directly
from the QCD Lagrangian, and confinement cannot even be proven.
Instead, one has to use "QCD-inspired", phenomenological models to
describe confinement.

\section{The QCD vacuum}

The expression for the strong coupling constant eq. (\ref{strong
force factor}) is derived from the full gluon propagator, $D(q^2)$
where $q^2=-Q^2$. This propagator is essentially the boson
propagator, but it has been modified due to vacuum polarization.

\bigskip

It is possible, to reach the same quantative conclusion involving
the QCD vacuum, which follows a line of reasoning by Gottfried and
Weisskopf \cite {weisskopf}.

\bigskip

Let us consider the empty vacuum, i.e., one where there are no
gluons. To this vacuum, add a pair of gluons of opposite colour
charges and spincomponents, and with an average separation $r$. The
energy for this pair is

\begin{equation}
E(r)=\frac A{4\pi r}-\frac{C\alpha _s}{4\pi r}\hspace{0.75cm}\mbox{for}%
\hspace{0.75cm} r\lesssim r_0,
\end{equation}
where A and C are constants. The first term is the kinetic energy
and the second one is the potential energy. $r_0$ is the distance
where $\alpha _s$ gets "large". For $r\ll r_0$, $\alpha _s$ is
small and $E(r)$ positive, but near $r_0$, $E(r)<0$! When $r>r_0$
the expression is no longer valid, but it should suffice to assume
that $E(r)$ increases with $r$.

\begin{figure}[h]
\begin{center}
\includegraphics{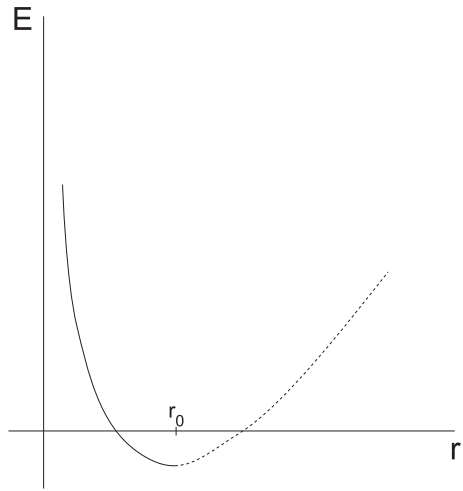}
\end{center}
\caption{\textit{The energy $E$ of the gluon pair versus distance $
r.$}}
\end{figure}

\bigskip

If this picture is correct, it seems to be energetically favourable
to create a gluon pair with opposite spins and colours out of the
''empty'' vacuum, to take advantage of the negative energy when
$r\sim r_0$.

\bigskip

The true vacuum can now be described as follows. The ''empty''
vacuum is unstable. There exists a state of lower energy that
consists of cells, each containing a gluon pair in a colour- and
spin-singlet state. The size of these cells is of order $r_{0}$.

\bigskip

This give one a crude description of how a colour-neutral
assembly of quarks is immersed into the gluon vacuum. The gluonic
cells will be displaced by the quarks, and therefore the quarks
will find themselves in a gluon-free ''bubble'' or ''bag''. A state
with quarks and gluons in the same bag has a higher energy, and
therefore the quarks "push away" the gluon field. The size of this
''bag'' is of the order of 1 fm. Since the true vacuum has an
energy density lower than the ''empty'' vacuum, the energy density
in the bag must be positive. Actually, it is proportional to the
bag volume, $V$:

\begin{equation}
E_B=BV. \label{Bag energy}
\end{equation}
$B$ is called the bag constant and has the dimension of
(energy)$^{4}$ provided that one sets $c=\hbar=1$ so that distances
are counted in 1/eV. The region outside the bag exerts a pressure
on the bag which is counteracted by the kinetic energy of the
quarks. It can be shown \cite {weisskopf} that this picture
describes the observed structure, size and low-lying spectra of
hadrons reasonably well if $B^{1/4}\approx 150$ MeV. The pressure
on the bag then amounts to $\sim 10^{23}$ atm.

\bigskip

Now one can compare the field of two opposite electric charges with
the field of two opposite colour charges at a separation larger
than $r_0$. In the electromagnetic case the field lines spread when
the separation grows.
\begin{figure}[h]
\begin{center}
\includegraphics{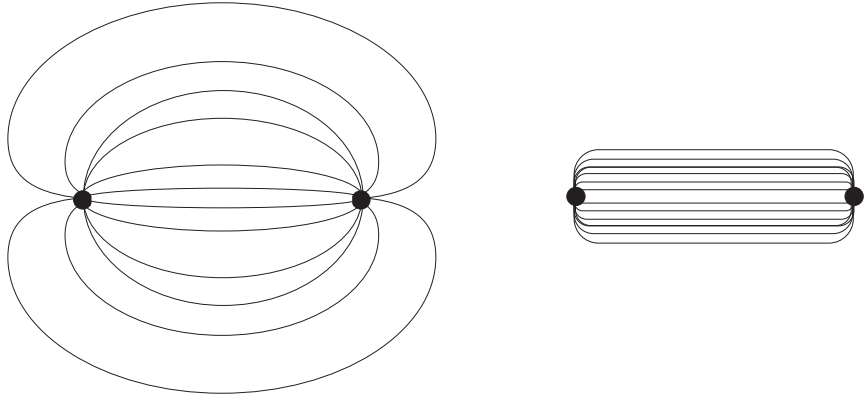}
\end{center}
\caption{\textit{Phenomelogical picture comparing the nature of the strong
force with the electromagnetical force. }}
\end{figure}
The number of electric field lines crossing a unit area decreases
like $r^{-2}$. In the colour case, the pressure of the true vacuum
compresses the field lines into a tube of diameter $r_0$. When
$r\gg r_0$ the number of field lines per unit area within this tube
remains constant, leading to a constant force, i.e., a linear
potential:

\begin{equation}
\phi (r)=ar\hspace{0.75cm}\mbox{for}\hspace{0.75cm}r\gg r_0.
\label{linear potential}
\end{equation}
The experimental value of the constant $a$ is $a\sim 0.6$ GeV/fm.

\bigskip

In conclusion, there is a constant force acting on a quark if one
tries to remove it from a hadron to distances larger than $r_0$. If
one would like to remove it completely one would need an infinite
amount of energy. However, long before that, the colour field would
"break up" into new quark-antiquark pairs.

\section{Bag models}

Since the potential in QCD grows indefinitely with distance,
coloured particles are confined to each other. The so-called bag
models have been invented to take this into account, because the
exact QCD formalism is virtually impossible to solve for bound
hadronic states (just like QED for atoms). In these models, quarks
are confined to a certain hadronic volume $V$. The Dirac equation
for the quarks becomes

\begin{eqnarray}
i\gamma ^\mu \partial _\mu \psi -m\psi=0 & \mbox{{\it inside}} \\
\psi=0 & \mbox{{\it outside}}. \nonumber
\end{eqnarray}
Then, the quark current through the surface is zero. This makes the
boundary condition read

\begin{equation}
n_\mu j^\mu =n_\mu (\overline{\psi }\gamma ^\mu \psi )=0.
\label{bag boundary condition}
\end{equation}

This bag model was proposed by Bogolioubov in 1967 \cite{Bo67}.

\section{The MIT bag model}

In the mid 1970s a group of physicists at the Massachusetts
Institute of Technology (MIT) showed that Bogolioubov's bag model
leads to energy-momentum conservation violation at the bag surface,
unless the internal pressure from the interior of the bag is
balanced by an external pressure. This led to a modified model, the
MIT bag model \cite{ch74,jo75,hk78}, where the true vacuum inside
the bag is partially destroyed by quarks carrying colour. This
mixture of quarks and true vacuum made it possible to treat the
physics of the interior of the bag by perturbative QCD, a so-called
perturbative vacuum inside the bag. This change in the bag model
leads to a new boundary condition. The requirement of pressure
balance at the surface is written as

\begin{equation}
-\frac 12n^\mu \partial _\mu (\sum_i\overline{\psi _i}\psi _i)|_S=B,
\label{MIT boundary condition}
\end{equation}
where $B$ is the bag constant, and the sum runs over all quarks
contained in the bag. For a spherical bag this condition is
equivalent to the requirement that the total energy contained in
the bag volume should be at a minimum with respect to the bag
radius $R$. Hence $\frac{\partial M}{\partial R}=0$, where

\begin{equation}
M(R)=\frac{(\sum_ix_i)}R+\frac{4\pi }3BR^3,
\label{Bag energy of radius}
\end{equation}
$x_{i}=\omega_{i}R$, and $\omega_{i}$ are the eigen-frequencies of
the solutions to the Dirac equation. The equilibrium radius is
obviously

\begin{equation}
R_0=\left( \frac{\sum_ix_i}{4\pi B}\right) ^{\frac 14}. \label{Eq
bag radius}
\end{equation}
It is possible to include real gluons in the bag even without
quarks. Such states, "glueballs" have not been found, although
signals of hybrid glueball/quark states are sometimes reported. The
appropriate boundary condition for the colour field is obtained
from the requirement that the colour-electric field should not be
able to penetrate outside the bag. The situation is analogous to
the boundary condition in classical electrodynamics for a medium
with $\varepsilon =0$ and $\mu =\infty $. By virtue of Gauss'
theorem an integration of the colour-electric field yields the
result that the total colour charge contained in the bag volume
vanishes.

\bigskip

There are several flaws in the MIT bag model, for instance,

\begin{itemize}
\item  that the mass of the pion comes out too large in all versions of the model

\item  that the boundary condition is not chirally invariant,
or equivalently, the MIT bag model violates PCAC, Partially
Conserved Axial Vector Current.
\end{itemize}

\bigskip

Several attempts have been made to overcome these difficulties. In
the chiral bag models a pion field has been added as an independent
degree of freedom \cite{BR79,TMT80}. In the cloudy-bag model
\cite{TMT80,TTM81}, or in the Tel Aviv model \cite{KE83}, the pion
cloud is allowed to penetrate into the bag. In these models, the
main goal has been to make the model chirally invariant at the
surface.

\bigskip

One of the models, the soliton bag model, where the large mass of a
quark outside the bag is generated by the coupling to a scalar
field, is very well suited for the description of the dynamical
properties of the bag, such as vibrations of the bag surface. This
model could be useful in the study of giant QGP bags.

\bigskip

It is important to remember that all the bag models are only
attempts to construct a model that explains the behaviour of the
particles as measured in experiments. As of this date, the
mechanism of confinement is well established experimentally, but
since perturbative QCD cannot be used in this region the
theoretical understanding of confinement is incomplete.

\chapter{The transition to a quark-gluon plasma}

\label{chaptrans}

\section{Hadrons as systems of quarks}

Ordinary matter is made out of leptons and hadrons. Hadrons are
strongly interacting particles consisting of quarks. We know from
statistical physics and thermodynamics that macroscopic systems of
ordinary matter exhibit changes in phases when the environment
changes in a specific way. Why should not nucleons, built of
quarks, exhibit some change in phase when the environment changes?

\bigskip

It is experimentally justified to describe collisions of nuclei,
i.e., systems of hadrons, in thermodynamic terms. The predicted
temperatures in the collisions are in accord with experimental
results. This temperature can be estimated from the energy
distribution of the emitted fragments. The only discrepancy appears
when the impact energy gets very high, the temperature does not
increase as fast as the model suggests. The temperature seems to
approach some kind of plateau when the collision energy grows. The
temperature of boiling water does not change during the phase
transition when the water enters its vapour phase. Would colliding
nuclei approach some kind of analogue phase change? It turns out
that the levelling-off of the temperature is due to the fact that
higher-mass hadrons are created in the collision. The energy in the
collision is high enough to convert into mass and create heavier
hadrons than protons and neutrons.

\bigskip

It is possible to look at such a creation of hadrons as a kind of
phase transition. It has a latent heat just as the liquid-vapour
transition in water. The temperature in the collision remains the
same until the most massive hadron have been created. Then the
temperature starts to increase again. Hagedorn has suggested that
there exists such a limit, and the limiting temperature should be
about\thinspace $1.5\cdot 10^{12}$ K \cite{highest mass hadron}.
This is not so far away from current accelerator experiments.

\section{From hadronic matter to QGP}

The ultimate phase transition would be the transition from hadronic
matter to a quark-gluon plasma. In the plasma phase the nucleons
and the higher mass hadrons created in the collision lose their
identity as individual particles.

\bigskip

It is important to remember the difference between the bonding of
an electron to a nucleus and the bonding between quarks in a
nucleon. It is possible with a finite investment of energy to
separate an electron from a nucleus, and turn a gas into a plasma
where all electrons move freely. But in a nucleon, one would need
an infinite amount of energy to separate a quark from its
environment.

\bigskip

There are two ways to create a plasma in a gas of atoms. One is to
heat the gas until the collisions between the atoms are violent
enough to rip away all electrons. Another is to compress the gas,
driving the atoms into close contact and make their electronic
shells overlap. Under these conditions the electrons are deconfined
and can move freely from one atom to the other.

\bigskip

As explained earlier, the quark interaction is negligible at short
distances (asymptotic freedom). On the other hand, when the
distance increases the interaction also increases (confinement).
These characteristics rule out the first way (heating) to create a
quark-gluon plasma, but not the second one (compression). Hence,
when making a quark-gluon plasma one does not need to set the
quarks free, only to push them into the same bag. It should be
added that it is at least theoretically possible to create a plasma
by heating, although the phase transition is not obtained by quark
deconfinement but by a more subtle mechanism \cite{chalmer
rapport}.

\section{The QCD phase transition}

At low energies, the QCD vacuum is characterized by nonvanishing
expectation values of vacuum condensates, the quark condensate having $<%
\overline{\psi }\psi >\approx (235$ MeV$)^3$ and the gluon condensate having
$<\alpha _sG_{\mu \nu }G^{\mu \nu }>\approx (500$ MeV$)^4$. The
quark condensate describes the density of quark-antiquark pairs
found in the QCD vacuum and is the expression of chiral symmetry
breaking\footnote{See Appendix A for an explanation of chirality
and chiral symmetry.}. The gluon condensate measures the density of
gluon pairs in the QCD vacuum and is a manifestation of the
breaking of scale invariance of QCD by quantum effects.

\bigskip

Broken symmetries in nature are often restored through phase
transformations at high temperatures. Some examples are
ferromagnetism, superconductivity and the transformation from solid
to liquid. In the quark-gluon plasma case, the phase transition
from hadronic matter to quark-gluon plasma would restore the broken
chiral symmetry. Nuclear matter at very high temperature would
exhibit neither confinement nor chiral symmetry breaking.

\bigskip

It is known from quantum field theory that chiral symmetry is
broken if the particles involved have non-zero mass. But in the QCD
vacuum, a non-zero value of the quark-antiquark condensate
expectation value $<\overline{\psi}%
\psi>$ has the same effect. The density of spin-0 quark-antiquark paris
has the chiral decomposition
$q\overline{q}=\overline{q}_Lq_R+q_L\overline{q}_R$, where
subscripts $R$ ("right-handed") and $L$ ("left-handed") denote
states with spin parallel and antiparallel to the direction of
motion. Hence the broken vacuum state contains pairs of quarks of
opposite chirality. If a left-handed quark annihilates a
left-handed antiquark, then the process is perceived as a change of
chirality of the free quark, exactly the same effect as a
non-vanishing quark mass. In reality though, the light quarks have
mass and the chirality of a light quark is not conserved, even if
the quark condensate vanishes.

\bigskip

When the temperature increases the interactions occur at shorter
distances, mainly by the weak coupling because the long range
strong interaction gets dynamically screened. Finite temperature
perturbation theory shows that the strong coupling constant $\alpha
_s$ falls logarithmically with increasing temperature \cite{ref 3,4
of search for}. The result that the quark-condensate parameter
vanishes at high temperature, makes it likely that the transition
between the low-temperature and the high-temperature manifestations
of QCD shows a discontinuity, i.e., a phase transition. The order of
this transition is believed to depend on the number of light, quark
flavours. Two massless flavours would generate a second-order
transition \cite{ref 6,7 search for} while three massless flavours
give a first-order transition.

\bigskip

Numerical simulations indicate that the transition temperature lies
in the range $150\pm 10$ MeV \cite{ref search for...}.

\chapter{The properties of a QGP}

\label{chapprop}

\section{Equation of state}
\label{seceqofstate}
Considering the energy scale in the hot QGP, i.e., $\sim 200$ MeV,
the approximation that the plasma only contains $u$ and $d$ quarks
which are massless should be quite reasonable. With this
approximation and the assumption that one can neglect all quark
interactions inside the plasma, one can derive the equation of
state \cite{QGP}.

\bigskip

First, one needs to count the degrees of freedom for the
constituents:
\begin{equation}
\mbox{Gluons:}\hspace{0.5cm}N_g=2(spin)\times 8(colour)=16
\label{Gluon degrees of freedom}
\end{equation}
\begin{equation}
\mbox{Quarks:}\hspace{0.5cm}N_q=2(spin)\times 2(flavour)\times 3(colour)=12.
\label{Quark degrees of freedom}
\end{equation}
The energy density residing in each degree of freedom is calculated
separately for the quarks and the gluons. The gluons form, without
the interactions, an ideal relativistic Bose gas of temperature $T$
which gives the energy density

\begin{equation}
E_g=\int \frac{d^3p}{(2\pi )^3}\frac p{(e^{\beta p}-1)}=\frac{\pi
^2T^4}{30},
\label{Gluon energy density}
\end{equation}
where $\beta=1/T$. For the quarks and the antiquarks one has to
introduce a chemical potential $\mu$ because, in general, there
will be a slight surplus of quarks over antiquarks in the QGP
created from normal atomic nuclei (heavy ions). At zero
temperature, $\mu $ is the energy needed to add another quark to
the plasma. Since no antiquarks are present at $T=0$ the energy
necessary to add an antiquark is zero. This does not imply that
$\mu=0$, because the additional antiquark may annihilate a quark
and release the energy $\mu $, assuming that the quark lies at the
surface of the Fermi sea. The chemical potential of the antiquarks
must therefore be chosen as $-\mu$.

\bigskip

The energy density for the quarks, treated as a relativistic Fermi
gas, cannot be calculated analytically if $\mu ,T\neq 0$. However,
the energy density for a quark and an antiquark is
\begin{eqnarray}
E_q & = & \int \frac{d^3p}{(2\pi )^3}\frac p{[e^{\beta (p-\mu
)}+1]}, \\ E_{\overline{q}} & = & \int \frac{d^3p}{(2\pi )^3}\frac
p{[e^{\beta (p+\mu )}+1]}, \\
\label{energysum}
E_q+E_{\overline{q}} & = & \int \frac{d^3p}{(2\pi )^3}\left\{\frac
p{[e^{\beta (p-\mu )}+1]}+\frac p{[e^{\beta (p+\mu )}+1]}\right\}
\\
 & = & \frac{7\pi ^2}{120}T^4+\frac{\mu ^2}4T^2+\frac{\mu ^4}{8\pi ^2}.
\nonumber
\end{eqnarray}
Considering baryon-number symmetric quark-gluon matter, i.e., $\mu
=0$, and multiplying with the respective degrees of freedom, the
energy density becomes
\begin{equation}
E=16E_g+12(E_q+E_{\overline{q}})=\frac{37\pi ^2}{30}T^4=\left(
\frac T{160\mbox{ MeV}}\right) ^4\left[ \mbox{GeV/fm}^3\right].
\label{QGP energy density}
\end{equation}
The expressions above use $k_{B}=c=\hbar=1$, giving $T$ the
dimension of energy, and length the dimension of (energy$^{-1}$).
The energy density inside a nucleon is four times the MIT bag
constant, $E_N=4B\approx 300-500$ MeV/fm$^3$, which together with a
transition temperature of $\sim 150$ MeV gives an energy density of
the QGP of $\sim 1-1.5$ GeV/fm$^3$. From this one can see that one
needs at least a factor of two higher energy densities than inside
a nucleon in order to make a QGP.

\bigskip

But what if $\mu \neq 0$? One can still use eq. (\ref{QGP energy
density}) but $\mu $ must now be computed, i.e., one has to know its
relation to the baryonic density $n_B$. This density is one third
of the density difference between the quarks and the antiquarks
multiplied with the degrees of freedom:
\begin{equation}
n_q=\int \frac{dp}{e^{\beta (p-\mu )}+1}, \label{Density of quarks}
\end{equation}
\begin{equation}
n_{\overline{q}}=\int \frac {dp}{e^{\beta (p+\mu )}+1},
\label{Density of antiquarks}
\end{equation}
\begin{displaymath}
n_q-n_{\overline{q}}=\frac \mu 6T^2+\frac{\mu ^3}{6\pi ^2}
\end{displaymath}
\begin{equation}
\label{Baryon density}
\Longrightarrow n_B=4(n_q-n_{\overline{q}})=\frac{2\mu }3T^2+\frac{2\mu ^3
}{3\pi ^2}.
\end{equation}
Hence $n_B=\frac 43\frac{\partial E}{\partial \mu }$ where
$E=E_{q}+E_{\bar{q}}$ is taken from eq. (\ref{energysum}). The
pressure and entropy of the plasma are given by
\begin{equation}
p=\frac 13E, \label{QGP pressure}
\end{equation}
\begin{equation}
s=\frac 13\frac{\partial E}{\partial T}. \label{QGP entropy}
\end{equation}
Under what conditions should the bag containing the plasma be
stable? A fair assumption would be that the external vacuum
pressure, characterized by the bag constant $B$, should not exceed
the internal pressure, i.e., $p=\frac E3\geq B$. The critical
temperature $T_c$ and the critical chemical potential $\mu _c$ can
be calculated using the eqs. (\ref{QGP energy density}) and
(\ref{QGP pressure}) evaluated at the critical pressure $p_c=B$:
\begin{equation}
B=p_{c}=T_c^4\left[ \frac{37\pi ^2}{90}+\left( \frac{\mu
_c}{T_c}\right)
^2+\frac 1{2\pi ^2}\left( \frac{\mu _c}{T_c}\right) ^4\right].
\label{Critical equation}
\end{equation}

\begin{figure}[h]
\begin{center}
\includegraphics{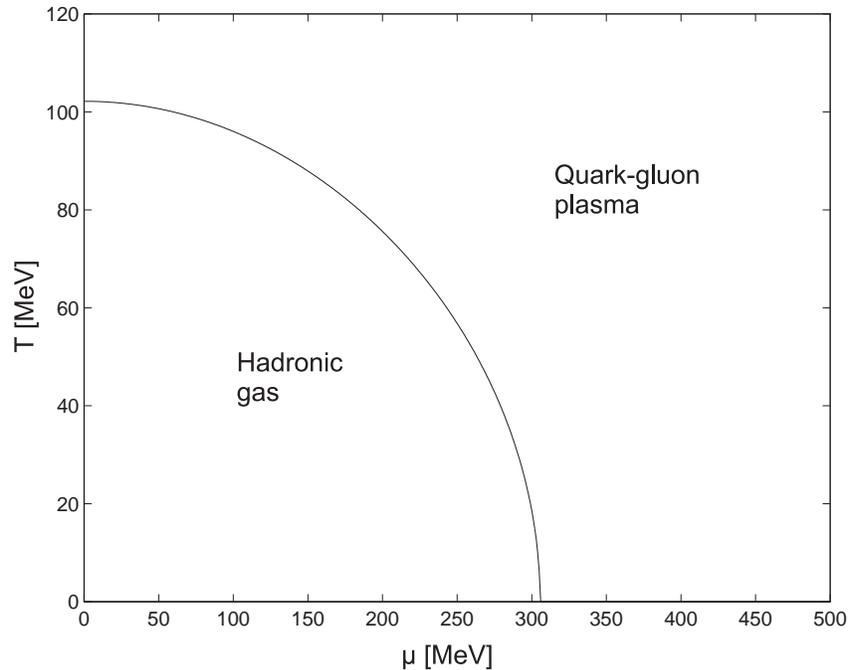}
\end{center}
\caption{\textit{A typical result from eq. (\ref{Critical equation}),
showing the relation between the critical values of the temperature
T and the chemical potential $\mu$. The bag pressure
$B^{1/4}=145$} MeV \textit{was used.}}
\end{figure}

\bigskip

Considering also perturbative interactions in the QGP, a
second-order thermal perturbative calculation gives \cite{QGP}
\begin{eqnarray}
E=\frac{8\pi ^2}{15}T^4\left( 1-\frac{15\alpha _s}{4\pi }\right)
+3\sum_f\left[ \frac 7{60}\pi ^2T^4\left( 1-\frac{50\alpha _s}{21\pi }%
\right) +\right. \\
\left. +\left( \frac 12T^2\mu _f^2+\frac 1{4\pi ^2}\mu _f^4\right) \left( 1-%
\frac{2\alpha _s}\pi \right) \right] +B,  \nonumber
\label{Energy density perturbation}
\end{eqnarray}
where $f$ is the quark flavour (neglecting quark masses). This
formula includes also the effect of strange quarks ($s\bar{s}$
pairs), provided they can be treated as massless. However, it does
not include short-range spin effects in QCD, so-called
colour-magnetic forces. These tend to favour quarks to form spin-0
pairs ("diquarks"). For a review of such effects, see
\cite{diquarkrev}. Would such effects be important, the QGP should
rather be treated as a boson gas, or a boson-fermion mixture. Such
non-perturbative phenomena cannot be rigorously treated within QCD.

\section{The formation of a QGP - experiments}

There are only two known situations where a QGP object could form

\begin{itemize}
\item  In a very hot and dense astrophysical environment, for instance, in
the early Universe or in the interior of a heavy neutron star or a
collapsing star, as will be discussed in a separate chapter

\item  In a high-energy heavy-ion collision taking place in an accelerator
or due to cosmic rays.
\end{itemize}
A central high-energy collision between heavy nuclei, resulting in
a compression of nuclear matter, is the most promising process fir
creating a QGP in a laboratory environment. There are presently
several such experiments running at the SPS collider at CERN and
the AGS collider at Brookhaven. Two more colliders are under
construction, LHC at CERN and RHIC at Brookhaven.

\[
\begin{tabular}{cccc}
\hline\hline
\textbf{Facility} & \textbf{Location} & \textbf{Starting date} & \textbf{%
Energy/nucleon} \\ \hline\hline AGS & Brookhaven, USA & 1986 & 4.84
GeV \\ SPS & CERN, Switzerland & 1986 & 17.2 GeV \\ RHIC &
Brookhaven, USA & 1999 & 200\ GeV \\ LHC & CERN, Switzerland &
$\sim $ 2005 & 5.4 TeV \\ \hline\hline
\end{tabular}
\]
There are three distinguishable processes in models of
ultra-relativistic heavy-ion collisions and QGP formation. The
first is the collision of the two nuclei and the formation of the
thermalized QGP. Several models have been developed to describe
this dynamical process, e.g. the partonic cascade model \cite{Ref
53 Ph and sign.} and the QCD string-breaking model based on the
Lund string model \cite{Ref 41 Ph and sign.}. An overview is given
in \cite {Ph and sign.}. The second process involves the
(relativistic) hydrodynamic behaviour of the plasma. The third
process is when the plasma has cooled to the critical temperature,
$T_c\approx 150-200$ MeV, and starts to hadronize.

\bigskip

Several detailed calculations have been presented \cite{Ref 7,62,65
Ph and sign.} in support of the view that a thermalized QGP will be
produced in the planned experiments at RHIC and LHC.

\section{The decay of a QGP}

The decay of a QGP is believed to take place along one of two
possible decay channels. It may expand until the density drops
below the threshold of stability for quark matter, or it may
hadronize through emission of particles, mainly pions, from the
surface of the confining bag.

\bigskip

It can be shown by a fairly simple reasoning \cite{QGP} that pion
evaporation cannot be the dominant process in the decay of the
plasma. There are three possible processes that can create pions on
the surface of a QGP bag:

\begin{itemize}
\item  A quark loses kinetic energy while bouncing off the surface from
inside.

\item  An antiquark loses kinetic energy while bouncing off the
surface.

\item  A quark and an antiquark annihilate at the surface, emitting a
pion.
\end{itemize}

The energy emission through these three processes is \cite{QGP}
\begin{equation}
\frac{d^3E}{dtd^2x}\approx 0.02\frac{T^6}{f_{\pi}^2}  \label{QGP decay rate}
\end{equation}
where $f_{\pi}$ is the decay constant of the pion and $x$ is a
spatial dimension. The cooling by pion emission increases with
temperature, but integrating eq. (\ref{QGP decay rate}), the pion
radiation falls below the thermal emission rate up to temperatures
of $300$ MeV. This leads to the conclusion that cooling by pion
radiation is probably not a major channel of energy loss.

\bigskip

This supports the other alternative, i.e., that the plasma will
expand until it reaches the critical temperature $T_c$ and then
convert into a hadronic gas, while maintaining thermal and chemical
equilibrium. A similar approach \cite{Ref 86 Ph and sign.} is that
the partonic reactions inside the bag change the parton density
locally so that a hadron can be formed. Hence, the reactions split
the bag into smaller regions, where the hadrons form due to local
density fluctuations.

\bigskip

The decay due to expansion and cooling hence seems to be the
dominant mechanism, and the lifetime of a QGP should be a few times
$10^{-23}$ s.

\chapter{Signatures of a QGP}

\label{chapsign} In order to detect a QGP one needs some clear experimental
signatures of its formation and/or decay. Many such signatures have
been suggested in the literature, and some of them are reviewed
here. Searching for a QGP formation is an event-by-event search
where the events detected by the detector can be connected to some
signature of a QGP. If the critical energy density is reached,
several types of reactions are possible, not only QGP formation. In
experiments one tries to select events with central collisions,
i.e., with a small impact parameter, where high mass densities are
most likely to occur. This can be done by focusing on events with,
e.g. high multiplicities, many nuclear fragments or protons or a
high symmetry around the beam axis.
\section{Kinematic probes of the equation of state}

Such probes are based on the energy density, pressure and entropy
of superdense hadronic matter as a function of the temperature and
chemical potential. One looks for a rapid rise in the effective
degrees of freedom, as expressed by the ratios $\frac E{T^4}$ or
$\frac s{T^3}$ over a small temperature range ($s$ being the
entropy). If a first-order phase transition takes place, these
quantities exhibit a discontinuity or a step-like rise \cite{Ref 17
Ph and sign}.

\bigskip

It is impossible to directly measure temperature and energy
density. Instead, one measures the average transverse momentum,
$<p_T>$, and the rapidity\footnote{ Rapidity, $y$, is a commonly
used generalization of the velocity
\par
\[
y\equiv \arctan h(v_{\parallel })=\arctan h(\frac{p_{\parallel
}}{p_0} )=\frac 12\ln \left( \frac{p_0+p_{\parallel
}}{p_0-p_{\parallel }}\right),
\]
where the momentum four-vector, $p^\mu
=(p_0,p_{\parallel },p_{\perp })$ and $p_{0}=E_{p}/c$. $E_{p}$ is the
energy of the incoming particle. The parallel direction
"$\parallel$" along the beam is taken as the spatial z axis, and
"$\perp$" is orthogonal to the beam. The rapidity is, unlike the
velocity, a relativistically additive variable for a change of
inertial system along the beam direction.} distribution of the
produced hadrons, $\frac{dN}{dy}$. These quantities can be
theoretically related to the variables $T,\,s$ and $E$. If a rapid
change occurs in the effective number of degrees of freedom, which
would happen if a QGP was formed, one would see an s-shaped curve
in a plot of $<p_T>$ versus $\frac{dN}{dy}$. This plot serves as a
phenomenological phase diagram, and expresses the simple fact that
a QGP decay is expected to give many hadrons, and that they would
get high $p_{T}$ values due to the high temperature.
\begin{figure}[h]
\begin{center}
\includegraphics{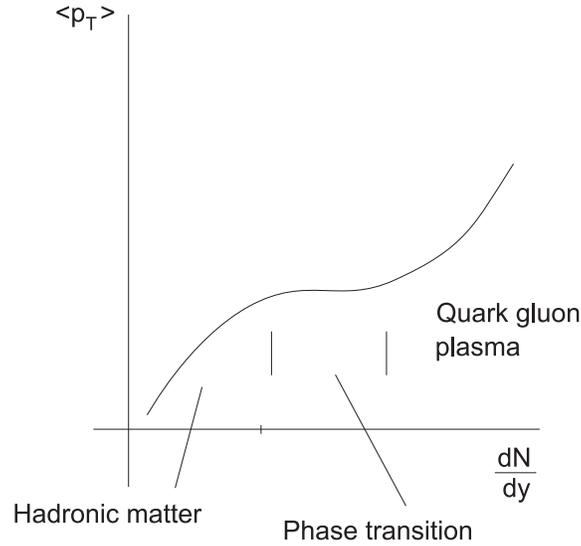}
\end{center}
\caption{\textit{A schematic plot of the expected behaviour of the mean
transverse momentum versus the density of produced particles in
rapidity space.}}
\end{figure}

There are other methods built on similar ideas, reviewed in
\cite{Ph and sign.}.

\section{Electromagnetic probes}

The clearest signal of the QGP is probably an excess of produced
lepton paris and photons. They probe the earliest and hottest phase
of the evolution of the QGP fireball without being affected by
final-state interactions. The drawback is that these signals are
supposed to be rather weak and cluttered with signals from hadronic
processes.

\bigskip

Lepton pairs have for long been considered clear probes of the QGP.
Thermal di-leptons are produced when a quark and an antiquark
annihilate in a QGP. However, this so-called Drell-Yan lepton-pair
production occurs also in normal hadron collisions. Recent
calculations \cite{Ref 243,244,245,246 Ph and sign.} show that
lepton pairs from the equilibrating QGP may dominate over Drell-Yan
production where invariant lepton masses are in the range $1-10$
GeV. If lepton pairs can be measured above the expected Drell-Yan
background from nucleon-nucleon collisions up to several GeV of
invariant mass, the early thermal evolution of the QGP phase can be
traced in a rather model independent way \cite{Ref 247 Ph and
sign.}.

\bigskip

Also photons from the QGP compete with the background of several
hadronic reactions. The hadronic radiation spectrum is not
concentrated in a single narrow resonance. In fact, a hadron gas
and a QGP in the vicinity of the transition temperature $T_c$ emit
photon spectra of roughly equal intensity and similar spectral
shape \cite{Ref 265 Ph and sign.}. If the QGP temperature lies
clearly above $T_c$, a signal of photons from the QGP phase would
be visible \cite{Ref 247 Ph and sign.,264,268 Ph and sign.}.


\section{Strangeness enhancement, related probes}

The most commonly proposed signatures of restored chiral symmetry
in dense baryon-rich hadronic matter are enhancements in
strangeness and antibaryon production. The production of hadrons
containing strange quarks is normally suppressed in nuclear
reactions \cite{Ref 136 search for}, because $s$ quarks do not
exist in the original heavy ions. When a QGP is formed, the
production of hadrons containing $s$ quarks is expected to be
saturated thanks to the production of $s\bar{s}$ pairs \cite{Ref
137 search for}. The yield of strange baryons and antibaryons is
therefore predicted to be strongly enhanced in the presence of a
QGP \cite{Ref 138,139 search for}.

\bigskip

There has been many observations of elevated strange-particle
production in nuclear collisions. However, such an enhancement
alone cannot prove the presence of a QGP. Strange particles can be
copiously produced in hadronic reactions, and several calculations
have been presented \cite{Ref 126,127 search for}. Nevertheless,
the enhancement of $\Lambda$ production over a wide rapidity range,
observed in heavy-ion collisions at 200 GeV/nucleon does not seem
to be explained by these models. Unfortunately, it would be
premature to conclude that a QGP has been formed. The QGP
calculations use the assumption that final-state interactions in
the decay of the QGP do not modify the hadron yield, while there
are strong reasons to believe that the strangeness-carrying hadrons
interact strongly. Therefore final-state interactions could very
well erase all traces of the plasma phase.

\bigskip

A strangeness abundance can still be a useful trigger for verifying
other signals of a QGP.

\bigskip

As stated earlier, a phase transition from hadronic matter to QGP
would restore the broken chiral symmetry, i.e., make the quarks
behave as if massless, or very light. This symmetry breaks down
again when the plasma decays into normal hadrons resulting in the
formation of so-called disoriented chiral condensates, DCC
\cite{Ref 145 search for}. The DCC would later decay into neutral
and charged pions, favouring a neutral-to-charge ratio different
from the value of 1/2 from isospin symmetry. Such events that
virtually violate isospin symmetry have been observed, namely the
so-called Centauro events from cosmic rays interacting in the
atmosphere \cite{Ref 150 Ph and sign.}.

\section{Hard QCD probes of deconfinement}

It is commonly believed that a $c\bar{c}$ pair produced by fusion
of two gluons cannot bind inside the QGP. Therefore, $J/\psi $
production is suppressed in the collision of two nuclei forming a
QGP compared to a normal hadronic process. This assumption is based
on the fact that a bound state cannot exist when the colour
screening length $\lambda_D\approx \frac 1{gT}$ is less than the
bound-state radius $<r_{J/\psi }^2>^{1/2}$ \cite{Ref 24 Ph and
sign.}. Computer simulations have shown that this seems to happen
slightly above the transition temperature $T_c$, say at $T\sim 1.2$
$T_c$. The formation of the $c\bar{c}$ system takes a time of about
1 fm/c, and therefore it could require a rather long QGP lifetime
before the suppression becomes effective. The details of modeling
$J/\psi $ suppression near $T_c$ are quite complicated and the
phenomenon has been extensively studied by several authors
\cite{Ref 184,185,186,187,188,189,190,191,192,127 search for,Ref
126}. There has been experimental observations of $J/\psi $
suppression in Pb-Pb collisions by the NA38 experiment at the
CERN-SPS \cite{Ref 207 Ph and sign.}. There are, however, other
hadronic mechanisms that could explain the results equally well
\cite {Ref 208 Ph and sign.}. Quarkonium suppression is
nevertheless believed to be the most promising signal so far of the
formation of a QGP.

\bigskip

Hard partons, i.e., quarks or gluons with very high energies, are
believed to be formed at the early phase transition. If such a
parton travels through a deconfined medium, it finds much harder
gluons to interact with than it would in a confined medium where
the gluons are constrained by the hadronic parton distribution
\cite{Ref Satz}. The mechanisms in this case are similar to those
responsible for the electromagnetic energy loss of a fast charged
particle in matter. Energy may be lost either by excitation of the
penetrated media or by radiation. Fast partons lose much more
energy per unit length in a QGP than in hadronic matter. Hence the
energy loss could tell whether or not the parton has travelled
through a QGP. On the other hand, the average transverse momentum,
$<p_{T}>$, of produced hadrons is supposed to be enhanced from a
QGP, due to the high temperature. A suppression of hard partons can
therefore be difficult to disentangle from the rise in $<p_{T}>$.

\section{Signatures - a summary}

It is obvious that a QGP can only be detected by the use of a
combination of signals from different stages of the
high-temperature phase of QCD. The signals discussed here all have
normal hadronic counterparts. However, the overall signals of a QGP
would presumably not be simultaneously duplicated by normal
hadronic reactions.

\bigskip

The QGP has yet to be uniquely observed or identified, but there
are
 data that appear as rather promising. One such example is
the detection of an enhancement of strange particles made at both
the AGS \cite{Ref 203 search for} and the SPS \cite{Ref
225,226,227,228,229 search for}. The enhancement in the $\Lambda $
yield measured over a large rapidity interval is difficult to
describe from purely hadronic interactions. An equilibrated QGP
with strangeness neutrality and large strangeness saturation, which
hadronizes and decays instantaneously, fits the SPS data \cite{Ref
141,236,237 search for}, but further experimental information is
necessary to rule out other explanations. The current (1996),
experimental situation is discussed in \cite{Ref search for}.

\chapter{Astrophysical aspects of QGP}

\label{chapastro} There are two situations in astrophysics and cosmology
where the QGP phase could be relevant. The first is inside compact
stars, i.e., neutron stars and, possibly, so-called quark stars. The
second situation is in the early Universe, before the quark-hadron
transition. There are other areas connected to those, which might
involve a QGP (one interesting example being during a supernova
collapse) but these are of a more speculative nature and will not
be dealt with here.

\bigskip

\section{QGP in the early Universe}

The early stages of the evolution of the Universe are described by
the standard Big Bang scenario. Its empirical basis are the
following observations:

\begin{itemize}
\item  The redshift in the spectra of distant galaxies due to the expansion
of the Universe.

\item  The existence of the $2.7$ K cosmic background radiation.

\item  The abundance of light elements, especially the He/H
ratio.

\item  An anisotropy in the background radiation corresponding to
a temperature fluctuation, $\Delta \mbox{T/T} \approx 6\cdot
10^{-6}$.
\end{itemize}
Indirect information about the early Universe is in principle
obtainable either through electromagnetic or gravitational
radiation, or, on the more speculative side, through exotic relics
like very massive particles or small black holes. In practice, the
only observations so far are from electromagnetic radiation. This
radiation gives information about the Universe when it was older
than $\sim 2\times 10^5$ years. This is due to the fact that at
that time the temperature of the Universe was somewhat lower than
the corresponding binding energies of electrons in light atoms. The
electrons and the light nuclei then formed stable atoms and the
Universe became transparent to photons. The cosmic background
radiation originates from this period.

\bigskip

The abundance of light elements supports the standard model of
primordial nucleosynthesis, i.e., the formation of light nuclei
directly after the Big Bang. This process took place when the
Universe was a few minutes old, and this is the earliest epoch of
which there is more or less certain information.

\bigskip

If one wants to study the very young Universe, i.e., when it was
younger than a few minutes, an extrapolation of the cosmological
model to earlier times is required. The basic physical theory used
in cosmology is Einstein's general relativity, but one also needs a
model for high-energy particle interactions.

\bigskip

Hadrons were formed in the quark-hadron transition when the
Universe was about $10^{-5}$ s old. The temperature was then about
$200$ MeV. Before this transition, all matter in the Universe was
contained in a small region with an enormous density, a QGP. The
transition is assumed to have been of first-order but this has not
been shown within QCD.

\bigskip

Further back in time, about $10^{-11}$ s after the Big Bang, the
electroweak phase transition took place. The temperature was then
about $100$ GeV, i.e., in the energy range of modern particle
accelerators. At $T\sim 100$ GeV there is a spontaneous symmetry
breaking\footnote{See Appendix B for the role of spontaneous
symmetry breaking in phase transitions.} of the electroweak
$SU(2)\times U(1)$ symmetry, through a weak first-order phase
transition. All leptons then acquire mass, while the intermediate
vector bosons split up into the massive $ W^{\pm }$ and $Z^0$ and
the massless photon.

\bigskip

Even earlier, at $t\sim 10^{-37}$ s, when $T\sim 10^{15}$ GeV,
another spontaneous symmetry breaking took place. Here, a unified
strong and electroweak interaction is believed to have been split
up by the Higgs boson.

\begin{figure}[h]
\begin{center}
\includegraphics{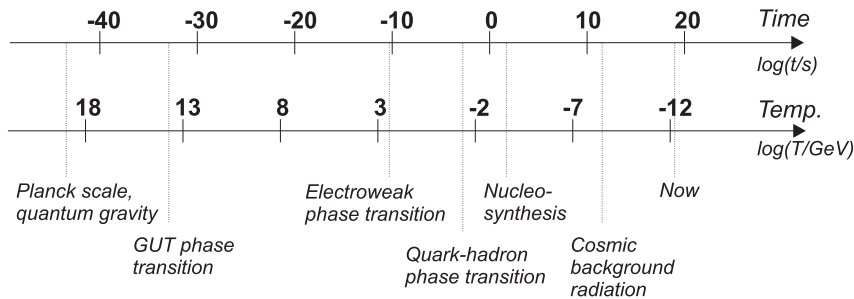}
\end{center}
\caption{\textit{Thermal history of the Universe.}}
\end{figure}

\bigskip

At $t\sim 10^{-43}$ s when the temperature was $T\sim 10^{19}$ GeV,
(the "Planck" temperature), quantum gravity effects were still
important. At this time, the cosmic horizon was about the size of
one particle, and since all mass/energy in the Universe was
contained in this small volume, gravity was important. Hence, a
theory of quantum gravity is needed to accurately describe this
period. All efforts to construct such a model have, however,
failed.

\bigskip

The fact that a quark-hadron transition took place in the early
Universe is quite undisputed, although one may dwell about whether
or not the quark phase was a QGP according to the standard particle
physics description. The main reason for the recent interest in
this transition is the implication the exact nature of the
transition could have on the abundance of light elements in the
Universe \cite{ref 42 schramm s 349}. However, it has also been
argued \cite{ref schramm} that the exact QGP behaviour is not so
crucial for the nucleosynthesis.

\bigskip

The actual transition is believed to have taken place by hadronic
bubble growth inside the QGP. This was suggested by Witten
\cite{ref Witten}, and although some of his conclusions have been
criticized \cite{critwitten1,critwitten2}, the bubble nucleation
mechanism is still in favour.

\bigskip

Witten suggested that at a temperature just below $T_c$, bubbles of
hadronic matter appear. Since this corresponds to a first-order
transition, a difference in energy density between the different
phases appeared; a so-called latent heat. As such bubbles expand
they expel heat, and a pressure equilibrium between the two phases
makes it possible for them to coexist.

\begin{figure}[h]
\begin{center}
\includegraphics{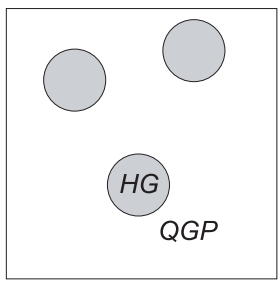}
\end{center}
\caption{\textit{Bubbles of hadronic gas (HG) begin to form inside the
QGP.}}
\end{figure}

\bigskip

When the Universe expanded the hadronic bubbles also expanded and
hence the formation of hadronic matter continued. When about 50\%
of the total volume had converted into hadronic matter, the QGP
phase began to form bubbles inside a hadronic sea.
\begin{figure}[h]
\begin{center}
\includegraphics{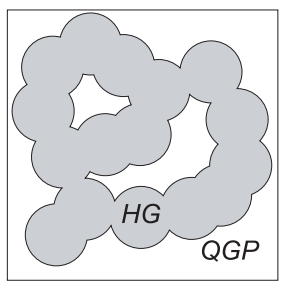}
\end{center}
\caption{\textit{Here more than 50\% of the total volume is occupied by hadronic
gas, and the QGP shrinks to bubbles.}}
\end{figure}
The further expansion of the Universe resulted in a loss of heat,
while the QGP bubbles shrank until they finally disappeared.
Thermal equilibrium seems to be a fair assumption for these rather
slow processes \cite {ref Witten}.
\begin{figure}[h]
\begin{center}
\includegraphics{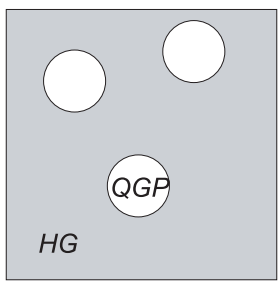}
\end{center}
\caption{\textit{The further expansion of the Universe made the QGP bubbles
shrink until they disappeared and led only the hadronic gas to
participate in the primordial nucleosynthesis.}}
\end{figure}
Witten also suggested the primordial QGP might have survived the
quark-hadron transition and constitute some of the dark matter in
the present Universe.

\bigskip

For a more thorough description of the present understanding of
quark-hadron transition in the early Universe, see \cite{ref
Ignatius}.

\section{QGP and compact stars}

This section reviews the possible existence of QGP in compact
stars, i.e., neutron stars and so-called quark stars. The material
is to a large extent taken from Glendenning's book "Compact Stars"
\cite{refcompstar}.

\bigskip

The birthplace of a star is a cloud of interstellar gas. The cloud
consists mostly of molecular hydrogen with varying density. The
temperature in the cloud varies from $10$ K to $2000$ K in
different spatial regions. When a local cluster of high-density
regions experiences a perturbation, e.g. a shock wave, it will
collapse gravitationally if the mass of the cloud is near the
critical "Jeans" mass. The compression continues until the internal
thermal pressure balances the gravitational contraction energy. The
core has then reached a high-enough temperature and density to fuse
hydrogen into helium. The contracted cloud has become a
main-sequence star.

\bigskip

The thermonuclear fusion continues even after all hydrogen in the
core has been used. Now helium burns into carbon, and the fusion
process continues until iron has been formed.

\bigskip

These processes do not depend on the mass of the star. The fusion
rate is faster for heavy stars, but the processes are the same for
all stars.

\bigskip

It is important to remember that fusion will start when and where
the temperature and density is high enough in the star. At the
beginning, fusion takes place only in the core, but the temperature
and density in the outer layers of the star may be high enough for
fusion to spread.

\bigskip

When fusion stops in the core of a light star of a few solar masses
M$_{\odot}$, the thermal pressure decreases and the star starts to
contract again. This increases the temperature and density all over
the star, making hydrogen burn in the outer layers of the star.
Such a star is called a
\textit{red giant}. All this repeats for helium, carbon etc. These processes
can occur in an explosive manner, ripping away most of the star to
form a planetary nebula. The remaining core of the star is too
light to maintain fusion and forms a
\textit{white dwarf} with a surface temperature of around $8000$ K.
It radiates energy for about 10$^{10}$ years and thereafter becomes
a \textit{black dwarf}.

\bigskip

For a heavier star (m $>8$ M$_{\odot }$) the evolution is somewhat
different from the period when iron has formed in the core. The
burning process then continues in the outer layers of the star,
adding to the iron content in the core. Gravity compresses the core
until the electrons are captured by the protons; a process called
inverse beta decay. The pressure in the core suddenly decreases and
an enormous implosion occurs, which raises the central temperature
in the core up to several tens of MeV ($\sim 10^{11}$ K). The core
neutronizes, i.e., turns electrons and protons into neutrons. It
decreases in volume, creating a shockwave that rips the outer
regions of the star apart and creates a supernova in the sky. The
total energy release is some 10$^{46}$ J. At this point the core
has reached its final equilibrium state, composed of neutrons,
protons, hyperons and leptons, and at the very centre, maybe a QGP.

\bigskip

If the mass of the star exceeds the so-called Oppenheimer-Volkoff
mass limit, where the internal thermal pressure cannot balance the
gravitational compression, a black hole is formed. There are other
types of collapse suggested for the formation of a black hole. See
\cite{refcompstar} for a review.

\bigskip

The density in the core of a neutron star is enormous ($\sim
10^{15}$ g/cm$^3$), which means that the energy density inside a
neutron star could be high enough for a QGP to form. Since the core
has a higher density than the outer regions, a QGP core should be
surrounded by a hadronic crust. Such stars are called hybrid stars.
In the extreme case of a so-called quark star there is no hadron
crust at all. For further reference, see chapter 8 in Glendenning's
book \cite{refcompstar}.

\bigskip

\begin{figure}[h]
\begin{center}
\includegraphics{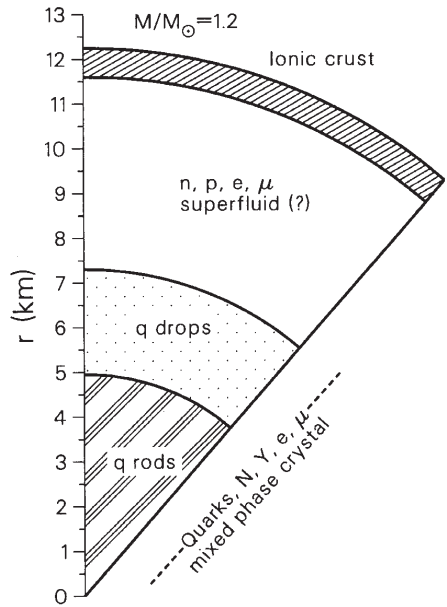}
\end{center}
\caption{\textit{A section through a neutron star that contains an
inner region of mixed phases including a crystalline lattice of
deconfined and confined matter and a surrounding liquid of neutrons
and leptons. A thin crust of heavy ions forms the
stellar surface \protect\cite{refcompstar}.}}
\label{hybintstruc}
\end{figure}

Two possibilities exist concerning the existence of a QGP inside a
hybrid star. It could have a pure QGP core with a distinct phase
boundary against the hadronic crust. But there could also be a
mixed phase with QGP and hadronic matter. Such a mixture could even
be in the form of a crystalline lattice of various geometries, with
the rarer phase immersed in the dominant one. Fig.
\ref{hybintstruc} shows the internal structure of a hybrid star.

\begin{figure}[h]
\begin{center}
\includegraphics{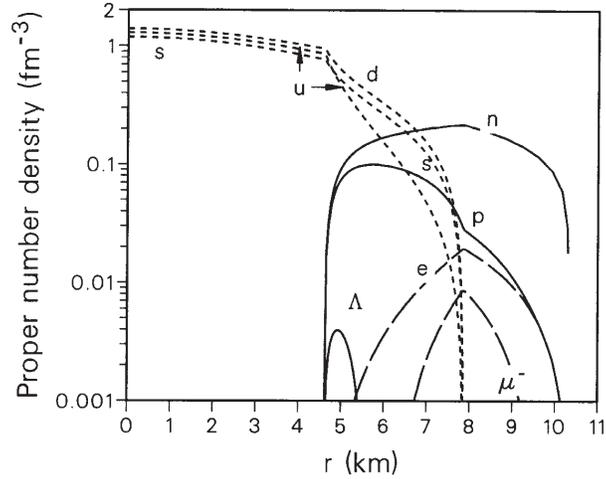}
\end{center}
\caption{\textit{Population densities per unit volume in a hybrid star with
mass 1.42 M$_{\odot }$ versus distance from the star centre
\protect\cite{refcompstar}.}}
\label{figstarpop}
\end{figure}

Is it possible to model the internal structure of a hybrid star?
With a bag constant $B^{1/4}=180$ MeV, and the coupling constants
lying in the range of accepted values, the population of quarks,
baryons and leptons will be as in Fig. \ref{figstarpop}.

\bigskip

A neutron star is normally too cold to be directly observed, with
one exception though; so-called pulsars. A pulsar is a highly
magnetized neutron star that rotates very quickly and thus emits
radiation along its axis of rotation, creating a pulsed signal with
a very stable period. This was actually discovered in 1968 by Bell
\cite{ref 125 com star}, and the total number of discovered pulsars
is now around 700. They are observed under various circumstances,
usually as isolated sources, but sometimes in binary orbit with
another white dwarf or neutron star.

\bigskip

Pulsars lie in the ms to s range with an average period of 0.7 s.
They are very stable and the fastest pulsar known, the PSR 1913+16,
has a period of $1.55780644887275\pm 0.00000000000003$ ms. The
accuracy here is due to terrestrial time standards. The pulsars are
therefore the most accurate clocks in the Universe. 

\bigskip

As one might guess, there is a physical limit of the rotational
velocity for a neutron star. When the centrifugal forces overcome
the gravitational contraction, the star will break up and emit
matter. For a neutron star of mass $1.44$ M$_{\odot }$ this limit
corresponds to a period of \thinspace $0.3$ ms (assuming that the
structure of the star is optimized to achieve maximum rotational
frequency).

\begin{figure}[h]
\begin{center}
\includegraphics{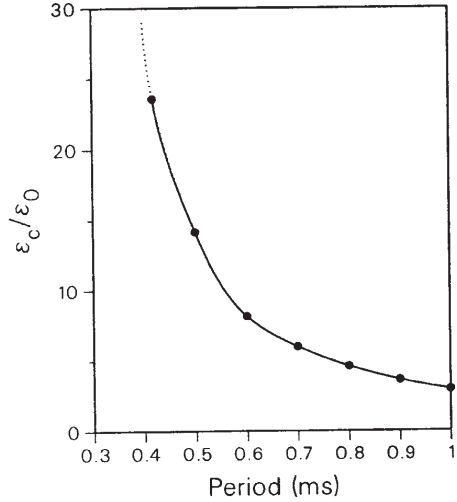}
\end{center}
\caption{\textit{The least possible theoretical central density of a
star with mass 1.44 M$_{\odot }$ required for stable rotation
versus the rotational period. The energy density $\epsilon_{c}$ is
normalized to $\epsilon_0$, the normal nuclear energy density
\protect\cite{refcompstar}.}}
\label{figstarrot}
\end{figure}

As can be seen in Fig. \ref{figstarrot}, the central density
required to achieve submillisecond periods grows rapidly with
decreasing period. One can thus conclude that matter under normal
nuclear density is not able to keep the star together if the
rotational period is of the order of $0.1$ ms. A reasonable
assumption is therefore that if a submillisecond pulsar is
discovered, its core must be denser than neutron matter, i.e.,
containing QGP in some form.

\section{QGP and dark matter}

The concept of cosmic dark matter is well established today. There
are clear signals from different observational sources for
discrepancies between the observed mass density and the dynamical
behaviour of galaxies. The luminous mass simply cannot account for
the observed rotations. This has been known for a long time, and
many models have been proposed in terms of non-luminous "dark"
matter, which ultimately could account for 90\% of the total mass
of the Universe.

\bigskip

One such model was proposed by Witten \cite{ref Witten} in 1984.
The dark matter was suggested to be leftover quark objects from the
quark-hadron transition in the early Universe. Witten showed that
under certain circumstances, QGP bags, or ''quark nuggets'' as he
called them, could survive to the present and form lumps of quark
matter of size $\sim 10$ cm with a density of $\sim10^{15}$
g/cm$^3$.

\bigskip

There are more recent, and detailed, analyses of Witten's idea.
Madsen argued \cite{Ref Madsen} that QGP bags with baryon number
A$<10^{35}$ cannot be dark-matter candidates. Alan, Raha and Sinha
\cite{Ref dm preprint} claimed that QGP bags with baryon number
10$^{39}\leq A\leq 10^{49}$ are cosmologically stable and could
solve the dark matter problem and even close the Universe
gravitationally.

\bigskip

Explaining dark matter with QGP objects is appealing in the sense
that it relies on the well-established idea that all nuclear matter
was once in the form of QGP.

\chapter{Stability of cosmic QGP objects}

\label{chapfound} \bigskip

\section{Introduction}

The second part of this thesis is an investigation of the
possibility of stable QGP bags. These bags are assumed to be
leftovers from the quark-hadron transition in the early Universe,
following Witten's idea \cite{ref Witten}. The basic idea is that
if a pressure equilibrium occurs at the edge of the QGP bag, the
bags could be stable. The dominant decay mechanism for a QGP is
cooling by expansion and then hadronization, but if a pressure
equilibrium would be present at the edge of the bag, no expansion
would take place and the QGP would be stable. This possibility
opens up new ways of explaining astrophysical phenomena such as
dark matter and gamma-ray bursts.

\bigskip

This investigation can be divided into two parts:
\begin{itemize}
\item  To determine the size where the gravitational pressure
balances the other pressures occurring at the edge of the bag.

\item  To determine the corresponding QGP mass and compare the result
with the Schwarzschild radius.
\end{itemize}
The calculations necessary to achieve these goals will be performed
in several different configurations, varying the equation of state
in different ways.

\bigskip

The question of quark matter stable against gravitational collapse
have been investigated before, in the sense of stable quark or
hybrid stars, e.g. \cite{horvath,qgpgrav indier}. Compared to those
models, the present work is built on similar equations of state,
but with a different formation mechanism, since it assumes that the QGP
objects are primordial.
\section{The model}

The model used for the stable QGP bag is simple. Four different
pressures exist:

\begin{itemize}
\item  A gravitational pressure.

\item  A vacuum pressure, characterized by the bag constant $B$.

\item  A kinetic pressure, from the motion of the quarks.

\item  A Fermi pressure of the quarks.
\end{itemize}

The kinetic pressure is identically zero if the temperature is
zero. If a stable solution is to be found some kind of repulsive
force must be taken into account. This is the degeneracy pressure
of the quarks. The total pressure in the QGP bag will be divided
into three parts:
\begin{equation}
p_t=p_{dg}+p_k-B.
\end{equation}
$B$ is the negative (inward acting) vacuum pressure and $p_k$ is
the outward kinetic pressure. $p_{dg}$ stands for the sum of the
gravitational and the degenerate (Fermi) pressure. The total
pressure $p_t$ will start at some value $p_c$ at the centre of the
bag and decline to zero at the edge of the bag.
\begin{figure}[h]
\begin{center}
\includegraphics{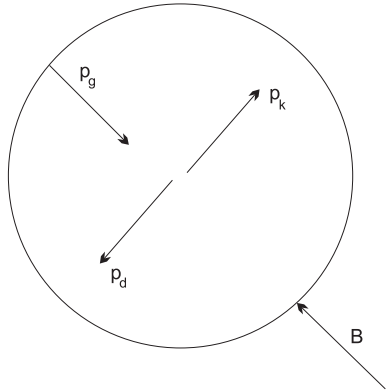}
\end{center}
\caption{\textit{Phenomenological picture of the pressures acting on the QGP
bag.}}
\end{figure}
The actual calculations rest upon certain approximations and
assumptions:

\begin{itemize}
\item  The quarks in the QGP bag are $u$, $d$ and $s$ quarks, all taken to be
massless.

\item  The plasma is taken to be a perfect fluid, i.e., the pressure is
isotropic in the rest frame of each fluid element so that shear
stress and heat transport can be neglected.

\item  The plasma is assumed to behave like a mixture of relativistic
boson and fermion gases (gluons and quarks).

\item  The quark-hadron transition in the early Universe was a first-order
transition. \

\item  The metric used to describe a static spherical star can also be used
to describe a QGP bag, i.e., the so-called
Tolman-Oppenheimer-Volkoff eq. (\ref{TOVeq}) is valid for a QGP
bag.

\item  One of the parameters needed to be set at each calculation is the
central pressure, $p_c$. The interesting range of this parameter is
assumed to begin at the pressure equivalent to normal nuclear
energy density, $\varepsilon_{nuc}\sim 170$ MeV/fm$^3$
(corresponding to $\sim 3\cdot 10^{14}$ g/cm$^3$).

\item  No decay process other than cooling by expansion and later
hadronization is taken into account.
\end{itemize}

\chapter{The crucial equations}

\label{chapgoveq}

\section{The Tolman-Oppenheimer-Volkoff equation}

The standard equation describing the interior of a compact star is
the Tolman-Oppenheimer-Volkoff (TOV) equation \cite{TOV}. It is
derived from the Einstein field equations with a static,
spherically symmetric metric governing the gravitational field for
a static spherical star. In addition, the mass-energy tensor for
the system is taken to represent a static perfect fluid. In the
original calculations, the energy density and the pressure were
taken to be zero outside the star. This leads to a boundary
matching where the mass, integrated from the centre of the star to
the edge, is set to be equal to the mass term in the external
Schwarzschild metric.

\bigskip

However, when applying this formalism to an MIT bag, with its
external vacuum pressure $B$, the energy density outside the bag
will be $4B$. This seemingly strange assumption, inherent to the
MIT bag model, does not significantly influence the metric.

\bigskip

The TOV equation is derived from the assumption of a static
spherical star taken as a perfect fluid. It is derived in many
textbooks in general relativity, e.g. in \cite{schutz}. It results
in the following set of equations:
\begin{equation}
\label{TOVeq}
\frac{dp}{dr}=-\frac{[\epsilon (r)+p(r)][m(r)+4\pi r^3p(r)]}{r[r-2m(r)]},
\end{equation}
\begin{equation}
\frac{dm(r)}{dr}=4\pi r^2\epsilon (r),\hspace{1cm}m(r)=4\pi
\int_0^r\epsilon (r^{\prime })r^{\prime }{}^2dr^{\prime },
\end{equation}
\begin{equation}
p(r=0)=p_c,
\end{equation}
\begin{equation}
p(r=R)=0.
\end{equation}
These equations are expressed in gravitational and natural units,
$c=\hbar=G=1$. $p$ is the total pressure, $\epsilon $ is the energy
density, $m$ is the mass and $r$ is the radial coordinate.

\bigskip

The TOV equation requires just a static spherical mass-energy
distribution and a perfect fluid. It does not depend on a certain
equation of state that relates energy density and pressure.

\section{The equation of state}

The equation of state for a non-interacting QGP consisting solely
by $u$ and $d$ quarks along with gluons was discussed in section
\ref{seceqofstate}. Allowing also $s$ quarks changes the equation
of state somewhat. The pressure is, as previously stated, divided
into four parts, where the kinetic part is derived for a mix
between three fermion gases and one boson gas. These circumstances
change the equation of state to
\begin{equation}
\epsilon (r)=3p_t(r)+4B.
\end{equation}
The kinetic pressure is now with three quark flavours $(f)$:
\begin{equation}
p_k=\frac{8\pi ^2}{45}T^4+\sum_f\left( \frac 7{60}\pi ^2T^4+\frac
12T^2\mu
_f^2+\frac 1{4\pi ^2}\mu _f^4\right),
\end{equation}
and the chemical potential $\mu _f$ is given for each quark
flavour. When taking the lowest-order gluon interaction into
account, the kinetic pressure changes into
\begin{eqnarray}
\label{pkekv}
p_k=\frac{8\pi ^2}{45}T^4\left( 1-\frac{15\alpha _s}{4\pi }\right)
+\sum_f\left[ \frac 7{60}\pi ^2T^4\left( 1-\frac{50\alpha _s}{21\pi
}\right) +\right. \\
\left. +\left( \frac 12T^2\mu _f^2+\frac 1{4\pi ^2}\mu _f^4\right) \left( 1-%
\frac{2\alpha _s}\pi \right) \right].   \nonumber
\end{eqnarray}
This expression has been derived in \cite{ref compstar 290,compstar
291,compstar 292}. Here one can obviously study also the case when
quarks do not interact, by setting $\alpha_{s}=0$.

\bigskip

After introducing $p_t$ into the TOV equation one gets
\begin{equation}
\frac{dp_{t}}{dr}=-\frac{\left(4p_t+4B\right) \left[ 4\pi \int_0^r\left(
3p_t+4B\right) r^{\prime }{}^2dr^{\prime }+4\pi r^3p_t\right]
}{r\left[ r-2\cdot 4\pi \int_0^r\left( 3p_t+4B\right) r^{\prime
}{}^2dr^{\prime }\right] } \label{TOV eq}
\end{equation}
This integro-differential equation cannot be solved analytically,
and basic numerical methods will be used in the following
calculations.

\chapter{The method}

\label{chapmethod}

\section{The algorithm}

Equation (\ref{TOV eq}) is solved with two simple numerical
algorithms. The derivative is changed according to
\begin{equation}
\frac{dp}{dr}\rightarrow \frac{p_{i+1}-p_i}h,  \label{derdiff eq}
\end{equation}
and the integral is changed into
\begin{eqnarray}
4\pi \int_0^r\left( 3p_t+4B\right) r^{\prime }{}^2dr^{\prime
}\rightarrow 4\pi \frac h2\cdot \left\{ [p_t(0)+4B]\cdot
0^2+2[p_t(h)+\right. \\
\left. +4B]\cdot h^2+2[p_t(2h)+4B]\cdot(2h)^2+...+[p_t(nh=r)+4B]\cdot r^2\right\},
\nonumber
\end{eqnarray}
where $h$ is the constant step length. rThese discretizations change
the TOV equation into
\begin{equation}
\frac{p_{i+1}-p_i}h=-\frac{\left( 4p_i+4B\right) \left( \sum_{int}+4\pi
r_i^3p_i\right) }{r_i\left( r_i-2\sum_{int}\right) },
\end{equation}
where the integral approximation is denoted $\sum_{int}$.

\bigskip

As can be seen, there is a singularity at $r_i=0$. This makes it
impossible to calculate $p(r=h)$ from eq. (\ref{TOV eq}). This
problem is not new, and methods have been developed to cope with
it. A polynomial expansion will be used for the first step, i.e., in
calculating $p(r=h)$, so that $p(r)=\sum_jp_jr^j$. Hence near
$r=0$,
\begin{equation}
p=p(\epsilon _c)+\left( p_c\Gamma _c/\epsilon _c\right) \left(
\epsilon
-\epsilon _c\right) +...\hspace{0.2cm} .
\end{equation}
$\Gamma _c$ is the "adiabatic index" $d[ln(p)]/d[ln(\epsilon)]$
evaluated at $\epsilon =\epsilon _c$, which also can be put on the
form $dp/d\epsilon $ taken at the same energy density $\epsilon
_c$. This expansion can now be written as
\begin{equation}
p(r)\approx p_c-2\pi \left( \epsilon _c+p_c\right) \left( p_c+\frac
13\epsilon _c\right) r^2+..., \label{sersol eq}
\end{equation}
so that $p(0)=p_{c}$, and $p(h)$ can be calculated. The resulting
formula
\begin{equation}
p_{i+1}=p_i-\frac{h\left( 4p_i+4B\right) \left( \sum_{int}+4\pi
r^3p_i\right) }{r_i\left( r_i-2\sum_{int}\right) }
\end{equation}
is to be iterated until the pressure $p_{t}=0$ is reached at some
radius $r$, containing the full mass $(\sum_{int})$.

\bigskip

A scaling procedure has been used. All involved quantities are in
km. This is possible by using gravitational and natural units:
\[
\begin{array}{ccc}
\left[ p\right] & = & 1/\mbox{km}^2, \\
\left[ \epsilon \right] & = & 1/\mbox{km}^2, \\
\left[ T\right] & = & 1/\sqrt{\mbox{km}}, \\
\left[ \mu _f\right] & = & 1/\sqrt{\mbox{km}}, \\
\left[ r\right] & = & \mbox{km}, \\
\left[ B\right] & = & 1/\mbox{km}^2, \\
\left[ m\right] & = & \mbox{km}.
\end{array}
\]
The following relationships hold:
\begin{equation}
B^{1/4},T,\mu _f : 1\mbox{ MeV}=1\cdot \left( \frac{2.6115\cdot
10^{-4}}{(197.33)^4}\right)^{1/4}=6.4421\cdot 10^{-4}\,
1/\sqrt{\mbox{km}},
\end{equation}
\begin{equation}
\epsilon ,p : 1\mbox{ MeV/fm}^3=1\cdot \frac{2.6115\cdot 10^{-4}}{197.33}=1.3234\cdot
10^{-6}\,1/\mbox{km}^2,
\end{equation}
\begin{equation}
m : 1\mbox{ kg}=1\cdot \frac{1.4766}{1.989\cdot
10^{30}}=7.4238\cdot 10^{-31}\mbox{ km}.
\end{equation}
When using the scaling expressed above, the normal nuclear density
$\varepsilon _{nuc}=3\cdot 10^{14}$ g/cm$^3$ $\approx 170$
MeV/fm$^3$ becomes $\epsilon_{nuc}\approx 2.2\cdot 10^{-4}$
1/km$^2$ and bag constant $B=(150)^4$ (MeV)$^4$ becomes $B\approx
8,7\cdot 10^{-5}$ 1/km$^2$.

\section{Numerical stability and errors}

The difference formula (\ref{derdiff eq}) for the derivative has an
error of order $O(h)$, while that of the integral approximation has
an error of order $O(h^2)$. It is in principle possible to reduce
the step size $h$, but it is then possible that one has to use the
series solution, eq. (\ref{sersol eq}), to determine the pressure
close to the singularity, i.e., more than one step out of $r=0$.

\bigskip

When it comes to the stability criterion of the iteration procedure
no theoretical examination has been made. The approach here has
been to reduce the step size by 50\% as long as significant
differences, ($>1$\%), remained between two iterations in the
mass/radius relationship for the QGP bag. The final step size was
$10$ m, i.e., $h=0.01$, which proved to be a suitable compromise
between accuracy and CPU time.

\section{Parameter variations}

In the calculations there are seven variable parameters; $B,\alpha
_s $, $T$, $\mu _u$, $\mu _d$, $\mu _s$ and $p_c$. The
central pressure has been varied between $1.7-1.3\cdot 10^{5}$
times the corresponding normal nuclear density.
The temperature has been varied between 0 and 2000 MeV, ($10^{12}$
K). No significant change in the bag size was caused by this
variation of temperature. The same is true for the chemical
potentials $\mu _u$, $\mu _d$ and $\mu _s$. The strength of the QCD
interaction, $\alpha_{s}$ has been varied between 0 and $2.5$,
which did \textbf{not} alter the bag size significantly. The only
important variation that did alter the bag size (apart from that of
the central pressure) was in the bag constant. The value of
$B^{1/4}$ was varied between 120 and 180 MeV.

\chapter{Results and discussion}

\label{chapresult}

\section{Results}
\begin{figure}[tbp]
\begin{center}
\includegraphics{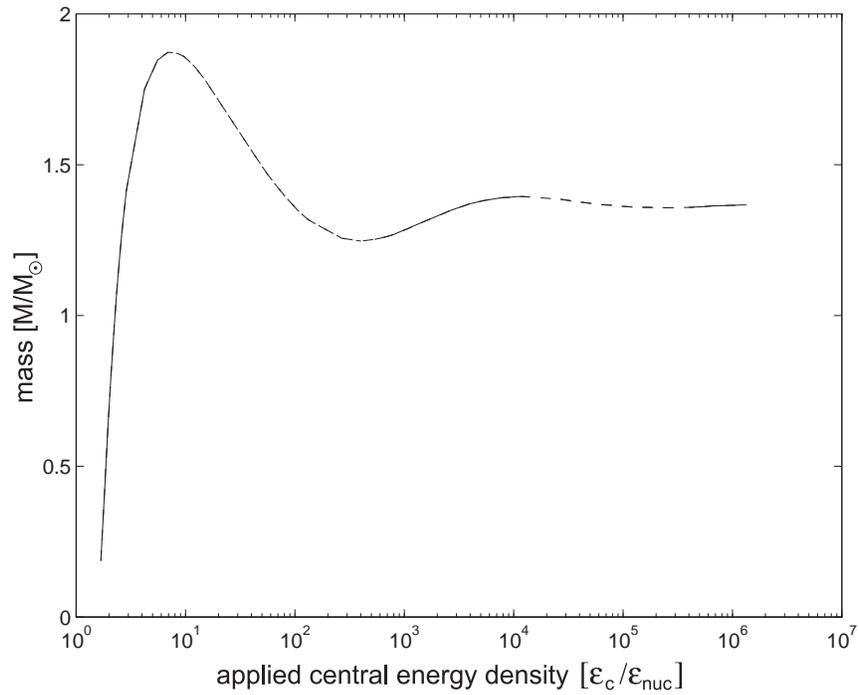}
\end{center}
\caption{\textit{Mass vs. applied central energy density. Stable
configurations occur only in the density intervals with the
full-drawn line. The parameter values are $ B^{1/4}=150$} MeV,
$T=0$ MeV, $\mu_u,\mu _d,\mu _s=0$ \textit{and} $\alpha_{s}=0$.}
\label{stabfig}
\end{figure}

When calculating the stable configurations for these QGP objects, a
stability criterion built into the TOV equation must be considered.
This is known as Le Chatelier's principle and can be formulated as
\cite{refcompstar}
\begin{equation}
\frac{\partial M(\epsilon_{c})}{\partial \epsilon_{c}}>0.
\label{lechat}
\end{equation}
When applying this condition to the results presented here, the
number of stable configurations is limited. As shown in Fig.
\ref{stabfig}, three regions in the range of the
applied central energy density $\epsilon_{c}$ allow stable
configurations.
The first region is only interesting in the upper part, since at
lower $\epsilon_{c}$, a QGP will not form. In the second region,
$\epsilon_{c}$ is far above the QGP transition allowing stable
configurations to occur. One should notice that at these
$\epsilon_{c}$ values, the chemical potential for the $c$ quark is
the same order of magnitude as its mass and therefore, the $c$
quark could be present in the centre of the QGP. This implication
is not included in the calculations. The small third region,
represents central densities about a thousand times higher than
those in the Universe at the time of the quark-hadron transition.

\bigskip

\begin{figure}[tbp]
\begin{center}
\includegraphics{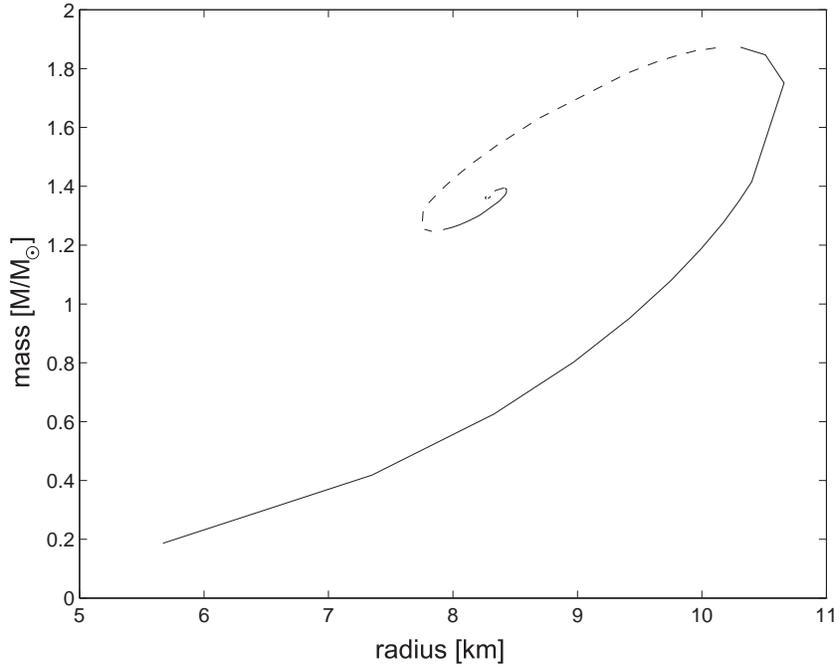}
\end{center}
\caption{\textit{This plot shows stable (full line), and
unstable (hatched line) configurations of the QGP bag. The
parameter values are $B^{1/4}=150$} MeV, $T=200$ MeV, $\mu _u,\mu
_d,\mu _s=0$ \textit{and $\alpha_{s}=0$. The central pressure is
varied from $10^{-5}$ to $100$} km$^{-2}$,
\textit{corresponding to a central energy density of $1.7$ to $1.3\cdot 10^{6}$
times $\varepsilon_{nuc}$.}}
\label{spicurve}
\end{figure}

Presenting the integration of the TOV equation in a mass-radius
diagram normally results in a spiral-like curve. This is true also for the
formalism presented here, as shown in Fig. \ref{spicurve}.

\bigskip
If a QGP bag is to be stable, i.e., to be able to
exist under a cosmological period of time, the temperature should be
zero so that no thermal photons are emitted. If the temperature is
set to zero in the calculations no visible change in the
mass-radius relationship occurs within the accuracy of the
iterative procedure, i.e., with a step in $r$ of 10 m.

\bigskip

\begin{figure}[tbp]
\begin{center}
\includegraphics{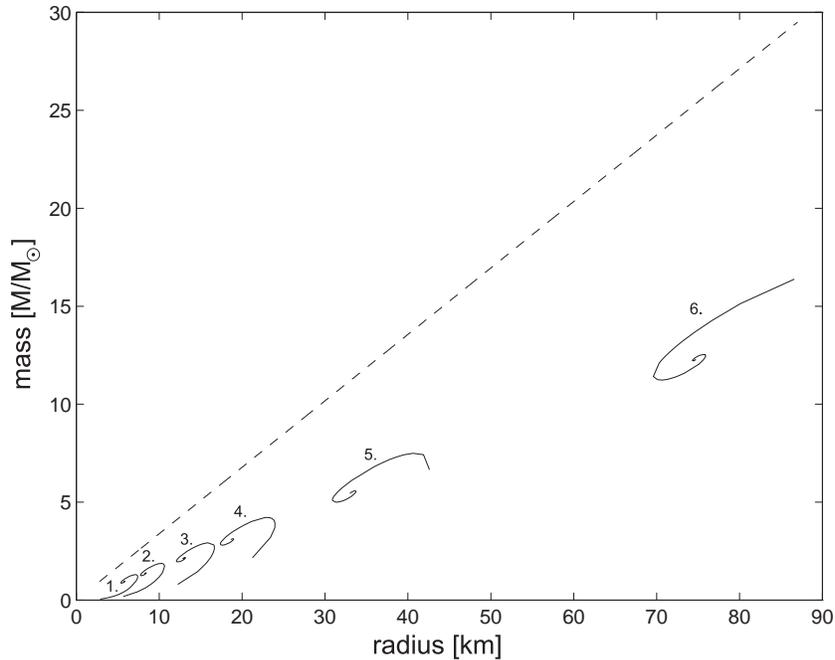}
\end{center}
\caption{\textit{The mass-radius relation for varying bag pressure
B (full line): B$^{1/4}$=180} MeV \textit{(curve 1.), 150} MeV
\textit{(2.), 120} MeV \textit{(3.), 100} MeV \textit{(4.), 75} MeV
\textit{(5.) and 50} MeV \textit{(6.). The hatched line corresponds
to the Schwarzschild limit. All other parameter values are the same
as in Fig. \ref{spicurve}.}}
\label{bvary}
\end{figure}

The eq. (\ref{pkekv}) for the kinetic pressure $p_k$ does
not depend much on the strong coupling constant $\alpha_{s}$, even
for $\alpha_{s}$ values up to 2. Consequently, the mass-radius
relationship is, in practice, independent of $\alpha_{s}$. The same
result is achieved when varying the chemical potentials of the
quarks. The only parameter variation that influences the
mass-radius relationship substantially is the bag constant as can
be seen in Fig. \ref{bvary}.

\bigskip

With the choice of parameter values in Fig. \ref{spicurve}, it is
obvious that one has two possible maxima regarding the mass of the
bag. One lies around $10.5$ km with a mass about 1.9 M$_{\odot }$.
This is achieved when the applied central energy density is about
10 $\epsilon_{nuc}$. The other lies at $8.5$ km with a mass about
1.4 M$_{\odot}$. Here, the applied central energy density is
roughly 10$^4$ $\epsilon_{nuc}$.

\bigskip

Using the basic formula for the Schwarzschild mass-radius
relationship, $r_s=2Gm/c^2$, it is improbable with a gravitational
collapse into a black hole, even at maximum mass configurations.
\begin{figure}[tbp]
\begin{center}
\includegraphics{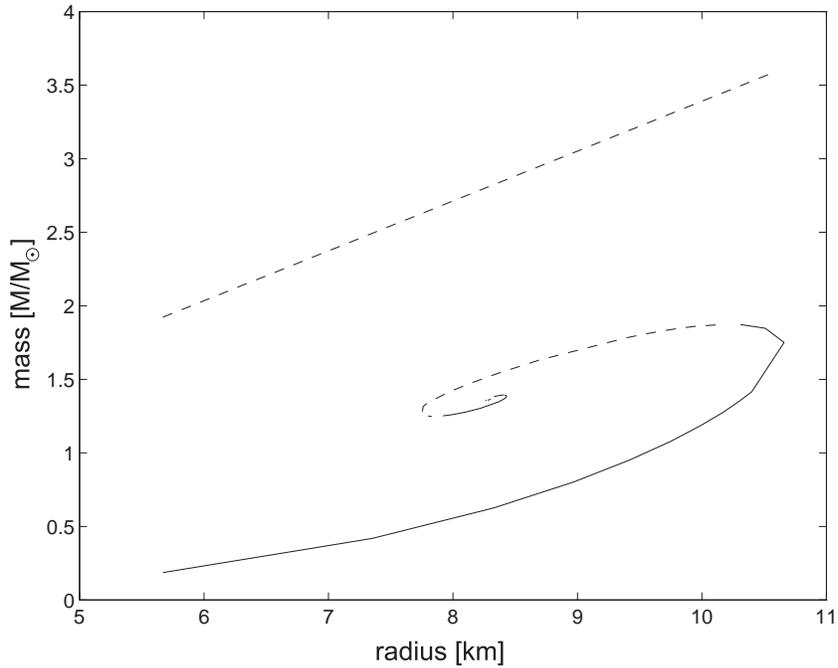}
\end{center}
\caption{\textit{The same results as in Fig. \ref{spicurve}
complemented with the mass-radius relation corresponding to the
Schwarzschild radius (upper hatched line).}}
\label{extschr}
\end{figure}
One might ask if there is any internal part that has a mass-radius
relationship near that given by the Schwarzschild relation. One
then needs to look at the energy density profile of the bag and
compare the integrated mass with the Schwarzschild mass. No such
situation appears with reasonable parameter values.

\bigskip

When varying the bag pressure $B$, as done in Fig. \ref{bvary}, the
regions in $\epsilon_{c}$ resulting in stable mass-radius
configurations is slightly shifted. One can also see, that
significant changes in $B$ affect the mass-radius configurations,
but not in a way that makes gravitational collapse more probable.
The value of $B$ is traditionally taken from quantum-mechanical
calculations involving the MIT bag model, where $B$ has been
adjusted to make the calculated hadron masses agree with
experimental data. In this thesis, the same values are used in a
vast extrapolation to macroscopic objects, which is customary in
astrophysical applications of the MIT bag model. However, one cannot be
certain that the value of $B$ is the same regardless of external
circumstances. In the calculations presented here, it is assumed
that $B$ does not alter with the outside metric. The possibility
that non-Euclidian geometry change the value of $B$ is not
examined.

\section{Discussion}

Based on these calculations, one can conclude that a QGP bag
obeying the chosen equation of state has a radius of either
7.5 - 8.5 km with mass 1.2 - 1.4 M$_\odot$ or 9 - 10 km with mass
1.6 - 1.8 M$_\odot$, while still being stable against evaporation and
gravitational collapse. This result agrees with other quark matter
stability calculations \cite{horvath,khadkikar}. The size and mass
of these bags make them interesting as dark matter candidates as
well as possible sources for gamma-ray bursts \cite{astropek}.

\bigskip

The calculations also show that the connection between
$\epsilon_{c}$ and the average energy density is rather weak, at
least when $\epsilon_{c}>100\,\epsilon_{nuc}$. At high
$\epsilon_{c}$, the energy density drops very fast with increasing
radius close to the centre and assumes a behaviour rather similar
to other, lower $\epsilon_{c}$ configurations, further out. This
fact could be due to numerical errors close to the centre and
remains to be investigated.

\bigskip

One circumstance that seriously could affect the possibility for
QGP bags to survive the quark-hadron transition is the cosmic
horizon size at the time of the transition, i.e., $10^{-5}$ s after
Big Bang. At this time, \cite{kolb and turner} states that the
radius of the Universe was about $10^{-11}$ times the present size,
i.e., about $10^{12}$ km.

\bigskip

The standard model in cosmology states that the average density in
the Universe at the transition was about $10^{20}$ kg/m$^3$. It is
possible that local density fluctuations could have created regions
that contained QGP at a considerably higher density than the
average at the time of the transition. These regions later expanded
with the Universe and possibly split up into smaller regions of QGP
when fluctuations inside the region started local hadronization
processes. Gravity retarded the expansion of these regions and the
size necessary for a stable QGP bag could have been achieved. This
process, when gravity counteracts the spatial expansion is a
dynamical process that would be very interesting to look further
into. Also, it would be of crucial importance to know if a QGP can
exist in the intermediate density region between $\epsilon_{nuc}$
and a few times $\epsilon_{nuc}$. This region is considered to lie
below the density needed to create a QGP out of compressed nuclei.
However, such investigations lie beyond the scope of the present
thesis.

\newpage {\LARGE \textbf{Acknowledgements}}
\addcontentsline{toc}{chapter}{Acknowledgements}

\bigskip

I wish to thank my supervisor, Professor Sverker Fredriksson at the
Division of Physics, Lule$\mbox{\aa}$ University of Technology for
encouragement, guiding and support, and Claes Uggla, Associate
Professor at the same division for introducing me to some of the
tools used in my calculations, especially in general relativity.
\\

\appendix
\newpage
\addcontentsline{toc}{chapter}{Appendix}
\chapter{Chiral symmetry}
This Appendix on chirality and chiral symmetry follows roughly
Donoghue, Golowich and Holstein \cite{Ref dyn std mod}.

\bigskip

If one lets $\psi (x)$ be a solution to the Dirac equation for a
massless particle:
\begin{equation}
i\not{\!\partial}\psi =0 \label{free dirac eq}
\end{equation}
and multiply this equation from the left by $\gamma _5$, one can
use the anticommutativity with $\gamma ^\mu $ to obtain
\begin{equation}
i\not{\!\partial}\gamma _5\psi =0. \label{no 2 free dirac eq}
\end{equation}
The following combinations can be formed:
\begin{equation}
\psi _L=\frac 12(1+\gamma _5)\psi, \hspace{1.5cm}\psi _R=\frac 12(1-\gamma
_5)\psi.  \label{phi combinations}
\end{equation}
The quantities $\psi _L$ and $\psi _R$ are solutions of fixed
chirality, in this case also called handedness. For a massive
particle moving with precise momentum, these solutions correspond
to the spin of the particle being, respectively, antialigned
(left-handed) and aligned (right-handed) relative to the momentum.

\bigskip

The matrices $\Gamma _{LR}=(1\pm \gamma _5)/2$ are
chirality projection operators, fulfilling

\begin{equation}
\begin{tabular}{c}
$\Gamma _L+\Gamma _R=1$ \\ $\Gamma _L\Gamma _L=\Gamma
_L,\hspace{1cm}\Gamma _R\Gamma _R=\Gamma _R,
\hspace{1cm}\Gamma _L\Gamma _R=\Gamma _R\Gamma _L=0 $.
\end{tabular}
\label{chiral identities}
\end{equation}
Chirality is a natural label to use when referring to massless
fermions thanks to its Lorentz invariance. A left-handed particle
is left-handed to all observers. A physical example of this is the
neutrino in the standard model. Neutrinos are left-handed chiral
particles, i.e., if they are massless. There is no evidence for
right-handed neutrinos.

\bigskip

Chirality is easy to add to the Lagrangian formalism. The
Lagrangian for a non-interacting massless fermion is
\begin{equation}
L=i\overline{\psi }\not{\!\partial}\psi \label{Lagrangian free
part}
\end{equation}
or
\begin{equation}
L=L_R+L_L=i\overline{\psi }_L\not{\!\partial}\psi
_L+i\overline{\psi }_R\not{\!\partial}\psi _R. \label{RL Lagrangian
free part}
\end{equation}
These Lagrangian densities are invariant under the global chiral
phase transformation
\begin{equation}
\psi _{L,R}(x)\rightarrow \exp (i\alpha _{L,R})\psi _{L,R}(x),
\label{chiral lagr transf}
\end{equation}
where the phases $\alpha _{L,R}$ are constant and real-valued.
Hence the Lagrangian density for massless particles has a chiral
symmetry. If we use
the famous Noether's theorem\footnote{%
Noether's theorem states that for any invariance of the action
under a continuous transformation of the fields in the Lagrangian,
there exists a classical charge $Q$, which is time independent and
is associated with a conserved current $J^\mu; \,\partial _\mu
J^\mu =0.$} we can associate conserved particle number current
densities
\begin{equation}
J_{L,R}^\mu =\overline{\psi }_{L,R}\gamma ^\mu \psi _{L,R}\hspace{1cm}%
(\partial _\mu J_{L,R}^\mu =0) \label{particle current dens}
\end{equation}
with this invariance. These chiral current densities form the basis
for the vector current $V^\mu (x)$:
\begin{equation}
V^\mu =J_L^\mu +J_R^\mu \label{vector current}
\end{equation}
and the axial vector current $A^\mu (x)$:
\begin{equation}
A^\mu =J_L^\mu -J_R^\mu. \label{axial vector current}
\end{equation}
It is now possible to construct conserved chiral charges:
\begin{equation}
Q_{L,R}(t)=\int d^3xJ_{L,R}^0(x), \label{chiral charge}
\end{equation}
and these represent the number operators for the chiral fields
$\psi _{L,R}$.

\bigskip

The vector charge $Q=Q_L+Q_R$ is the total number operator; whereas
the axial vector charge, $Q_5=Q_L-Q_R$, is the number operator of
the difference above. These charges simply count the sum and the
difference of the left-handed and the right-handed particles.

\bigskip

The $u$ and the $d$ quark have masses far below the QCD scale
parameter $\Lambda \simeq 0.2$ GeV which makes the assumption that
$m_u\approx m_d\approx 0$ quite reasonable. It is even a fairly
good approximation to treat the $s$ quark as massless, at least if
the temperature (in energy units) is comparable to the mass of the
$s$ quark. Due to the connection between mass and chiral symmetry
shown in eqs. (\ref{free dirac eq}-\ref{phi combinations}), chiral
symmetry is not present as long as the fermions have mass.

\bigskip

When referring to QGP, I have previously stated that chiral
symmetry is restored in the phase transition to QGP. This is only
approximately so because the quarks inside the plasma are not
massless. However, since the temperature and the scale parameter
$\Lambda $ is roughly $200$ MeV, the assumption that $m_u\approx
m_d\approx m_s\approx 0$ is fairly reasonable. The true chiral
symmetry restoration also depends on the fact that the quark
condensate expectation value vanishes at the phase transition. With
these approximations, chiral symmetry is said to be restored in the
QGP phase.

\chapter{The physics of phase transitions}

This description of the role of symmetry-breaking in phase
transitions is inspired by the book ''Cosmology'' by Coles and
Lucchin \cite {ref cosmology}.

\bigskip

The transition from some disordered phase to an ordered phase is
labelled by a lowering of the symmetry in the system, which
can be characterized by an order parameter $\Phi $. $\Phi $ is zero
in the most disordered phase and rises when more ordered phases
appear.

\bigskip

It is illustrative to take the solid/liquid phase transition as an
example. The ordered phase has a low degree of (discrete)
rotational symmetry, while the liquid gets an almost perfect
rotational symmetry when the atoms move around at random. The order parameter $%
\Phi $ in this case represents the deviation of the spatial distribution of
ions from the homogeneous distribution they have at $T>T_C$, the
melting point. If $T>T_C,\,\Phi =0$ and if $T<T_C,\,\Phi >0$.

\bigskip

When $\Phi $ grows significantly, a symmetry-breaking transition
has occurred. Such a transition can be caused either by an external
influence, such as heating, or by a change of the system itself.
Transitions due to external influences are called "induced", while
those due to internal phenomena are "spontaneous". For the
spontaneous symmetry-breaking processes it is convenient to
consider the free energy of the system, $F=U-TS$, where $U$ is the
internal energy, $T$ is the temperature and $S$ is the entropy. A
condition for the system to be in an equilibrium state is that $F$
is at a minimum. At $T=0$ the free energy coincides with the
internal energy. At $T>0$, whatever the form of $U$, an increase in
$S$ leads to a decrease in $F$, and is therefore energetically
favoured.

\bigskip

Phase transitions can be of two kinds; first- or second-order. If
$\Phi$ rises continuously from zero when $T<T_C$ (the transition
temperature), the transition is said to be of second-order, or
continuous. If $\Phi$ jumps discontinuously to a non-zero value
just below $T_C$ then the transition is of first order. Here, the
entropy also exhibits a discontinuity at $T_C$. In a first-order
transition, heat is adsorbed by the system in going from the
low-temperature to the high-temperature phase. This heat is the
latent heat, $Q_L=T_C\Delta S$, of the transition, where $\Delta S$
is the entropy change.

\begin{figure}[h]
\begin{center}
\includegraphics{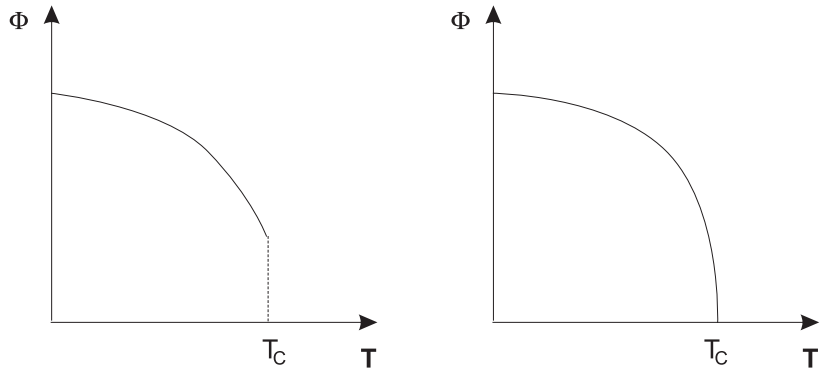}
\end{center}
\caption{\textit{Order parameter vs. temperature for a first-order
(left) and a second-order (right) transition.}}
\end{figure}

\bigskip

The QGP phase transition involves a spontaneous breaking of the
chiral symmetry, i.e., chiral symmetry is broken in hadronic matter,
but restored in the QGP. The order of the QGP phase transition has
not been uniquely determined. Two massless flavours would generate
a second-order transition \cite{ref 6,7 search for}, while three
massless flavours would give a first-order transition. In
astrophysical calculations that involve a QGP phase transition, one
often assumes that the transition is of first order.

\end{document}